\documentstyle[epsfig,longtable,graphicx]{mn}

\newcommand{\chem}[2] {$\rm{}^{#2}\kern-0.8pt#1$}

\newcommand{\barL}{\mbox{$\bar{L}$}}

\newcommand{\barH}{\mbox{$\bar{H}$}}

\newcommand{\barKs}{\mbox{$\bar{K_s}$}}

\newcommand{\barJ}{\mbox{$\bar{J}$}}

\newif\ifAMStwofonts

\ifoldfss

  \renewcommand{\chem}[2] {$\rm{}^{#2}\kern-0.8pt#1$}
  \ifCUPmtlplainloaded \else
    \NewTextAlphabet{textbfit} {cmbxti10} {}
    \NewTextAlphabet{textbfss} {cmssbx10} {}
    \NewMathAlphabet{mathbfit} {cmbxti10} {} 
    \NewMathAlphabet{mathbfss} {cmssbx10} {} 
  \fi
  \ifAMStwofonts
    \ifCUPmtlplainloaded \else
      \NewSymbolFont{upmath} {eurm10}
      \NewSymbolFont{AMSa} {msam10}
      \NewMathSymbol{\upi}     {0}{upmath}{19}
      \NewMathSymbol{\umu}     {0}{upmath}{16}
      \NewMathSymbol{\upartial}{0}{upmath}{40}
      \NewMathSymbol{\leqslant}{3}{AMSa}{36}
      \NewMathSymbol{\geqslant}{3}{AMSa}{3E}

       \let\le=\leqslant
       
    \fi
  \fi
\fi 

\ifnfssone
  \newmathalphabet{\mathit}
  \addtoversion{normal}{\mathit}{cmr}{m}{it}
  \addtoversion{bold}{\mathit}{cmr}{bx}{it}

 \newmathalphabet{\mathbfit} 
  \addtoversion{normal}{\mathbfit}{cmr}{bx}{it}
  \addtoversion{bold}{\mathbfit}{cmr}{bx}{it}
  \newmathalphabet{\mathbfss} 
  \addtoversion{normal}{\mathbfss}{cmss}{bx}{n}
  \addtoversion{bold}{\mathbfss}{cmss}{bx}{n}
  \ifAMStwofonts
    \ifCUPmtlplainloaded \else
      \UseAMStwoboldmath
      \makeatletter
      \new@mathgroup\upmath@group
      \define@mathgroup\mv@normal\upmath@group{eur}{m}{n}
      \define@mathgroup\mv@bold\upmath@group{eur}{b}{n}
      \edef\UPM{\hexnumber\upmath@group}
      \new@mathgroup\amsa@group
      \define@mathgroup\mv@normal\amsa@group{msa}{m}{n}
      \define@mathgroup\mv@bold\amsa@group{msa}{m}{n}
      \edef\AMSa{\hexnumber\amsa@group}
      \makeatother
      \mathchardef\upi="0\UPM19
      \mathchardef\umu="0\UPM16
      \mathchardef\upartial="0\UPM40
      \mathchardef\leqslant="3\AMSa36
      \mathchardef\geqslant="3\AMSa3E

       \let\le=\leqslant

    \fi
  \fi
\fi 

\ifnfsstwo
  \DeclareMathAlphabet{\mathbfit}{OT1}{cmr}{bx}{it}
  \SetMathAlphabet\mathbfit{bold}{OT1}{cmr}{bx}{it}
  \DeclareMathAlphabet{\mathbfss}{OT1}{cmss}{bx}{n}
  \SetMathAlphabet\mathbfss{bold}{OT1}{cmss}{bx}{n}
  \ifAMStwofonts
    \ifCUPmtlplainloaded \else
      \DeclareSymbolFont{UPM}{U}{eur}{m}{n}
      \SetSymbolFont{UPM}{bold}{U}{eur}{b}{n}
      \DeclareSymbolFont{AMSa}{U}{msa}{m}{n}
      \DeclareMathSymbol{\upi}{0}{UPM}{"19}
      \DeclareMathSymbol{\umu}{0}{UPM}{"16}
      \DeclareMathSymbol{\upartial}{0}{UPM}{"40}
      \DeclareMathSymbol{\leqslant}{3}{AMSa}{"36}
      \DeclareMathSymbol{\geqslant}{3}{AMSa}{"3E}

       \let\le=\leqslant

    \fi
  \fi
\fi 

\ifCUPmtlplainloaded \else
  \ifAMStwofonts \else 
    \def\upi{\pi}
    \def\umu{\mu}
    \def\upartial{\partial}
  \fi
\fi

\title{New Near-Infrared Surface Brightness Fluctuation Models}
\author[Mouhcine et al.]
       {M. Mouhcine$^{1,2}$, 
        R.A. Gonz\'alez$^3$,   
        M.C. Liu$^{4}$ \\
       $^1$ School of Physics and Astronomy, University of Nottingham, 
       Nottingham NG7 2RD\\
       $^2$ Observatoire Astronomique de Strasbourg (UMR 7550),
       11, rue de l'Universit\'e, 67000 Strasbourg, France \\
       $^3$ Centro de Radioastronom\'ia y Astrof\'isica, Universidad  
       Nacional Aut\'onoma de M\'exico, Campus Morelia, Michoac\'an 
       CP 58190, Mexico\\
       $^4$ Institute for Astronomy, University of Hawaii, 2680
       Woodlawn Drive, Honolulu, HI 96822}    
\date{Accepted ?.
      Received ?;
      in original form ?}

\pubyear{2001}

\begin{document}

\maketitle

\label{firstpage}

\begin{abstract}

We present new theoretical models for surface brightness fluctuations 
in the near-infrared. We show the time evolution of near-infrared
brightness fluctuation properties over large age and metallicity 
ranges, i.e., from 12~Myr to 16~Gyr, and from ${\rm Z/Z_{\odot}=1/50}$ 
to ${\rm Z/Z_{\odot}=2.5}$, for single age, single metallicity stellar 
populations. All the stellar models are followed from the zero age 
main sequence to the central carbon ignition for massive stars, or 
to the end of the thermally pulsing regime of the asymptotic giant 
branch phase for low and intermediate mass stars.

The new models are compared with observed near-infrared fluctuation 
absolute magnitudes and colours for a sample of Magellanic Cloud star 
clusters and Fornax Cluster galaxies. For star clusters younger 
than ${\rm \sim\,3\,Gyr}$, the predicted near-infrared fluctuation 
properties are in a satisfactory agreement with observed ones over
a wide range of stellar population metallicities. However, for older 
star clusters, the agreement between the observed and predicted 
near-IR brightness fluctuations depends on how the surface brightness 
absolute magnitudes are estimated. The computed set of models are not 
able to match the observed near-IR fluctuation absolute magnitudes 
and colours simultaneously. We argue that the observed discrepancies 
between the predicted and observed properties of old MC superclusters 
are more likely due to observational reasons.
\end{abstract}


\section{Introduction}
\label{intro.sec}

Over the last years, surface brightness fluctuations (SBFs) have been 
recognized as a powerful tool for providing accurate information on 
galaxy properties (Tonry, Ajhar, \& Luppino 1990, Tonry \& Schneider 
1988, Blakeslee et al. 2001). SBFs constitute a reliable technique to 
determine cosmic distances to galaxies, out to galaxy clusters such as 
Fornax or Coma with typical uncertainties of the order of 10\% or less 
(e.g. Tonry et al. 1990, Tonry et al. 1997, Liu \& Graham 2001, Mei et 
al. 2001). In fact, using near-infrared (near-IR) data obtained by the 
Hubble Space Telescope, Jensen et al. (2001, 2003) have shown that the 
SBF technique can determine distances reliably out to 
${\rm \sim\,150\,Mpc}$ (${\rm H_{\circ}=75\,km\,s^{-1}\,Mpc^{-1}}$). 
The usage of SBFs as a distance indicator requires that the bright 
end of the stellar luminosity function in galaxies is universal, 
or that the variation from galaxy to galaxy can be calibrated.

The problem of estimating the age and metallicity of stellar populations 
in galaxies is a highly controversial topic. The sensitivity of SBFs 
to stellar age and metallicity offers an opportunity to investigate 
the unresolved stellar content of galaxies, and thus provide new clues 
into their star formation history as a complementary tool to canonical 
methods based on integrated magnitudes and colours,
and spectral features. SBFs are a well-defined characteristic of 
any stellar system. As a consequence of the dependence of the SBFs on the 
second moment of the stellar luminosity function, they supply additional 
constraints on the properties of stellar systems that are not retrievable 
from the integrated luminosity. Near-IR SBFs 
are sensitive to the presence of bright and cool stars, and may therefore 
be used to detect the presence of intermediate age stellar populations in 
galaxies. In contrast with the optical SBFs, near-IR SBF magnitudes show 
a complex sensitivity to age and metallicity (Worthey 1993). The combination 
of near-IR SBF magnitudes and optical/near-IR integrated colours is suspected 
to lift significantly the age-metallicity degeneracy (Blakeslee et al. 2001; 
Liu et al. 2000).   

Extensive observational and theoretical work has been conducted to measure 
and to model the SBFs in the optical (e.g., Tonry et al. 1990, Worthey 1993, 
Liu et al. 2000, Blakeslee et al. 2001, Cantiello et al. 2003). However, 
the situation is completely different in the near-IR, whose observational 
windows have been, until recently, poorly exploited (Liu et al. 2002, 
Jansen et al. 2003). The scarcity of detailed near-IR SBF studies is due 
to the lack of both accurate empirical calibration and self-consistent 
models in the near-IR.
The near-IR SBF signal is more sensitive to the presence of late-type 
giant stars than the integrated stellar population luminosity, i.e., 
the brightest few magnitudes of the stellar luminosity function dominate 
the near-IR SBF signal (Liu et al. 2000). So, the accuracy of near-IR SBF 
predictions depends strongly on the reliability of the stellar evolution 
ingredients used to model late stellar evolutionary phases. 
The inability of stellar population models to account correctly for stellar 
systems near-IR properties is due to (i) the lack of a reliable understanding 
of late stellar evolution phases, and (ii) difficulties in the modeling of 
broad molecular bands that dominate the near-IR spectra of late-type stars. 

Mouhcine and collaborators have studied the spectrophotometric properties 
of stellar populations focusing on the most controversial predictions of 
stellar population synthesis models, i.e., the near-IR properties of 
intermediate age stellar populations (Mouhcine \& Lan\c{c}on 2002, 
Mouhcine et al. 2002, Mouhcine 2002, Mouhcine \& Lan\c{c}on 2003).
The purpose of these models was to provide a global framework for the 
study of various observational aspects of evolved stars in intermediate 
age populations, ranging from their contribution to the integrated light to 
the statistical weights of their carbon-rich and oxygen-rich representatives.
The theoretical predictions of the stellar population evolutionary models
were extensively compared to observational constraints. The models were
successful in reproducing a large variety of those constraints. The main 
step forward with these models is a better inclusion of the evolution of 
low and intermediate mass stars through a double-shell burning phase, 
known as the Thermally Pulsing Asymptotic Giant Branch (TP-AGB; see also 
Marigo et al. 2003 for another attempt to include TP-AGB in stellar 
population models).

Recently, Gonz\'alez et al.\ (2004) have presented the near-infrared SBF
measurements for a sample of Magellanic Clouds (MC hereafter) star clusters 
using the Second Incremental and All Sky Data releases of the Two Micron 
All Sky Survey (2MASS). This work opens up an opportunity of sharpening 
the capabilities of SBF measurements as diagnostics of unresolved stellar 
populations.  
The MC star clusters cover a wide range of ages, i.e., from a few Myr 
to more than 10~Gyr; on the other hand, they cover a narrow range of 
stellar metallicities, i.e., from ${\rm Z/Z_{\odot}\approx\,1/30}$ to 
${\rm Z/Z_{\odot}\approx\,0.5}$. These properties of the clusters, 
combined with the predicted sensitivity of near-IR SBFs to age and
metallicity, offer a unique way to disentangle the effects of age and 
metallicity on stellar population properties. On the other hand, the 
sensitivity of near-IR SBF signal to the details of stellar evolution 
provides an opportunity to constrain the accuracy of different ingredients 
of stellar population synthesis models. In this paper, we present new 
models of near-IR SBFs over a wide range of ages and metallicities for 
single-metallicity, single burst stellar populations, based on the stellar 
populations synthesis models of Mouhcine \& Lan\c{c}on (2002). 

The layout of this paper is as follows. In \S~\ref{sbf_nir}, we outline the 
ingredients of the stellar population synthesis models. In \S~\ref{comp} we 
discuss the evolution of absolute near-IR fluctuation magnitudes for a large 
range of ages and metallicities, and compare the theoretical predictions to 
observed near-IR SBFs. Finally, in \S~\ref{concl} the results of the present 
work are summarised.

\section{Near-Infrared Surface Brightness Models}
\label{sbf_nir}

The near-IR SBF models presented in this paper are based on the stellar 
population synthesis code described and tested observationally in 
Mouhcine \& Lan\c{c}on (2002, 2003). The code is designed to reproduce
the near-infrared properties of both resolved and unresolved stellar
populations, with an emphasis on intermediate age stellar populations.
In this section, we describe briefly the main ingredients of the stellar
population synthesis models, referring to the quoted papers for more 
details.

\subsection{Single Stellar Population Models}
\label{ssp}

The library of evolutionary tracks used in the population synthesis 
models is based on the models of Bressan et al. (1993) and Fagotto et al. 
(1994\,a,b,c). We will refer to these sets as the Padova tracks hereafter.
We include tracks with metallicities ranging from 1/50 to 2.5 times solar
metallicity. The sets of tracks cover major stellar evolutionary phases: 
from the main sequence to the end of the early-AGB phase of low- and
intermediate-mass stars, and to the central carbon ignition for massive
stars. As pointed out earlier, the contribution of intermediate mass 
stars to near-IR SBFs is large. So, the accurate inclusion of the 
latest evolutionary phase of low and intermediate mass stars, through 
the luminous double-shell burning regime at the end of the AGB, i.e. 
TP-AGB, is crucial. The Padova tracks do not extend to the end of the
TP-AGB phase. The extension of these tracks to cover the TP-AGB phase 
is needed. The evolution of low- and intermediate-mass through the 
TP-AGB evolutionary phase is followed using the so-called synthetic 
evolution modelling (Iben \& Truran 1978; Renzini \& Voli 1981). 
The inclusion of different processes affecting this stellar evolutionary 
phase (i.e., envelope burning, superwind, the third dredge-up) has been 
revealed to be crucial in order to match a variety of observational 
constraints (see Mouhcine \& Lan\c{c}on 2002). Only  a few of the widely 
used single metallicity, single burst stellar population, i.e., simple 
stellar population (SSP hereafter) models, account for these processes 
with modern prescriptions that take into consideration the complex 
interplay between them (Marigo et al. 2003). 

The initial conditions needed to calculate the evolutionary tracks of 
TP-AGB stars, i.e., the total mass (M), core mass (M$_c$), effective 
temperature (T$_{\rm eff}$), luminosity (L), and envelope chemical 
abundances, are taken directly from Padova tracks for different masses 
and metallicities. The properties of TP-AGB stars are then allowed to 
evolve according to semi-analytical prescriptions (Wagenhuber \& 
Groenewegen 1998). These prescriptions 
take into account the effect of metallicity on the instantaneous properties 
of TP-AGB stars. The evolution of TP-AGB stars is stopped at the end of 
the AGB phase since their contribution to the optical/near-IR light 
is almost negligible once they evolve behind this phase. The calculations 
of stellar evolution through the TP-AGB phase are performed using a mixing 
length parameter $\alpha\,=\,2$, Bl\"ocker's (1995) mass loss prescription 
($\dot{M}\,=\,6.13\,10^{-4}\eta\,L^{4.2}T_{eff}^{-2}M^{-3.1}$, with a 
mass loss efficiency $\eta\,=\,0.1$), dredge-up efficiency $\lambda=0.75$, 
and the critical core mass to trigger the third dredge-up 
${\rm M_c=0.58\,M_{\odot}}$; such is a set of stellar parameters and 
prescriptions that reproduces fairly well various constraints on TP-AGB 
stars in the solar neighbourhood and nearby resolved galaxies. 

A new set of stellar evolutionary models has been published recently 
by the Padova group that include several physical updates (Girardi et al. 
2000). We have relied, however, on the Padova 1994 evolutionary tracks 
for the following reasons. The population synthesis code we are using 
to compute near-IR SBFs was built initially to study the properties of 
intermediate age populations, and thus to constrain the free parameters 
of TP-AGB star models (see Mouhcine \& Lan\c{c}on 2002 for more details). 
This was done using old Padova tracks as initial conditions to the 
TP-AGB evolution models. The derived values of the free parameters of 
TP-AGB evolution models might not be valid if different evolutionary 
tracks and initial conditions for TP-AGB star models are used; in this 
case, the agreement between predicted and observed intermediate age 
stellar populations cannot be guaranteed. On the other hand, the old 
Padova evolutionary tracks appear to be more consistent with 
observational constraints than the new set of Padova tracks (e.g., 
Bruzual \& Charlot 2003; see also Gonz\'alez et al. 2004).

Stellar libraries are needed to assign spectral energy distributions 
to the stars of a synthetic population. For stars evolving through 
evolutionary phases other than the TP-AGB, we have used the theoretical 
stellar atmospheres of Kurucz (see Kurucz, 1979), Fluks et al. (1994)
and Bessell et al. (1989, 1991), as collected and re-calibrated by 
Lejeune et al. (1997, 1998). This library has the advantage of covering 
a broad range of temperatures, gravities and metallicities.

The Basel group has published recently a semi-empirical stellar 
spectral library that is calibrated to non-solar metallicities 
(Westera et al. 2002). Nevertheless, we have used the library of 
Lejeune et al. (1997, 1998), rather than that of Westera et al. 
(2002), for reasons that have to do with the available calibration 
data. Given the available calibration data, i.e., globular cluster
colour-magnitude diagrams and empirical colour-effective temperature 
relations, Westera et al. (2002) have concluded that it is impossible 
to establish a unique calibration of UBVRIJHKL stellar colours in 
terms of stellar metallicity that is consistent simultaneously with 
all the available data sets. On the other hand, the data that would 
allow a consistent calibration of the relationship between stellar 
near-IR colours and effective temperatures are extremely scarce. 
More data has to be collected to improve the quality of the 
calibration in the near-IR (see Westera et al. 2002 for a detailed 
discussion). On the other hand, Bruzual \& Charlot (2003) have shown 
that a combination of the Padova 1994 stellar evolutionary tracks 
with the Westera et al. (2002) library provides satisfactory fits 
to observed colour-magnitude diagrams of star clusters of different 
ages and metallicities. However, the comparison was performed mainly 
in optical bands. In addition, the evolution of near-IR fluctuation 
colours for old stellar populations as predicted by Gonz\'alez et al. 
(2004), who used the Westera et al. (2002) library, shows a behaviour 
which is not entirely consistent with the well established fact that 
near-IR colours of RGB stars get redder as their initial metallicity 
increases (see Sect.\,\ref{sbf_age} for more details). Theoretical and 
observational work is needed in order to understand if the discrepancies 
observed by Westera et al. (2002) are due to the colour-temperature 
relations, to the isochrones, or both.

The spectrophotometric properties of TP-AGB stars are different from 
those of giant stars. Their extended and cool atmospheres lead to the
formation of deep and specific spectral absorption bands. 
The photometric properties of oxygen-rich and carbon-rich stars 
TP-AGB stars are different; as an example, carbon stars show redder 
(H-K) colours than oxygen-rich TP-AGB stars at a given (J-H) colour. 
The dichotomy between carbon-rich and oxygen-rich TP-AGB stars is 
rarely considered in SSP models available in the literature. 
To account for TP-AGB star properties, we used the empirical library 
of average TP-AGB star spectra of Lan\c{c}on \& Mouhcine (2002). 

It is extremely difficult to estimate the effective temperature 
and the metallicity of variable TP-AGB stars. The uncertainties in 
current models of both stellar interiors and atmospheres of TP-AGB 
stars make it difficult to estimate reliably the fundamental physical 
parameters of these stars. Optical/near-infrared colours and narrow 
band indices are usually used as effective temperature indicators 
for M type TP-AGB stars. Note that narrow band indices show a larger 
scatter than optical/near-infrared colours due to the sensitivity 
of molecular bands to the structure of the outer atmosphere, which 
is affected by pulsation (Bessell et al. 1989; Alvarez \& Plez 1998; 
Lan\c{c}on \& Wood 2000). The effects of metallicity on the JHK 
spectra of M type TP-AGB stars are unclear. At a low spectral 
resolution, the shapes of JHK spectral features of stars in the Large 
and Small Magellanic Clouds (Lan\c{c}on \& Wood 2000) and near the 
Galactic Centre (Schultheis et al 2002) show no sizeable systematic 
differences with thOse observed for stars in the Solar Neighbourhood. 
However, the L-band spectral features of M type TP-AGB stars seem 
to depend on the metallicity (Matasuura et al. 2005). 

For M type TP-AGB spectra, two different effective temperature 
scales could be used. The first one assumes that variable TP-AGB 
stars may have the same effective temperature scale as static giant 
stars; the second one considers that the effective temperatures of 
variable TP-AGBs may differ from those of static giants. The shapes 
of the spectral features of stellar populations dominated by 
oxygen-rich stars depend strongly on the effective temperature scale 
of these stars. However, the adopted temperature scale has almost no 
effect on the evolution of integrated broad-band photometry of stellar 
populations (see Mouhcine et al. 2002 for a detailed discussion). 
The effect of metallicity on oxygen-rich TP-AGB star spectra, at a 
given effective temperature, is taken into account to first order, 
by using a metallicity-dependent effective temperature scale, i.e.,  
the relationship between effective temperature and (I-K) of Bessell 
et al. (1991). 

Theoretical modelling of carbon star spectra shows that the spectral 
features of carbon stars depend, in addition to the effective 
temperature and metallicity, on the C/O ratio, the 
$\rm{}^{13}{\kern-0.8pt}C$/$\rm{}^{12}{\kern-0.8pt}C$ ratio, and the
microturbulence (Gautshy-Loidl 2001). On the other hand, the C/O ratio 
depends on the metallicity and the evolutionary status of the stars 
along the TP-AGB sequence (Marigo 2002; Mouhcine \& Lan\c{c}on 2003). 
Despite the relative success of synthetic spectra to account for the 
shape of spectral energy distributions of carbon stars, we note that 
the effects of different parameters on different spectral features 
are degenerate, making it difficult to derive fundamental stellar 
parameters that are consistent from the stellar evolution point of 
view (see Loidl et al. 2001, and Lancon \& Mouhcine 2002 for a 
detailed discussion). Observationally, the JHKL spectrophotometric 
properties of carbon stars show complicated behaviour, i.e., some
spectral indices, such as the (J-H) colour and the HCN equivalent 
width, decrease with metallicity, while others, such as the (H-K) 
colour and the ${\rm C_{2}H_{2}}$ equivalent width, cover at low 
metallicities similar ranges than what is observed for carbon stars 
in the Solar Neighbourhood (\"Ostlin \& Mouhcine 2005, Matasuura 
et al. 2005). Similarly to what was done for oxygen-rich TP-AGB 
star spectra, the effect of metallicity on carbon star spectra is 
taken into account to first order, by using a metallicity-dependent 
effective temperature scale, i.e., the relationship between effective 
temperature and (R-H) of Loidl et al. (2001).

Once the physical ingredients describing different stellar evolutionary
phases, and the recipe used to estimate the magnitude and colours of the 
stars are specified, the only free parameters of a SSP model are those 
that describe the initial mass function (IMF). For the rest of the paper
we use the standard Salpeter (1955) IMF. The adopted lower and upper 
mass cut-offs are M$_{\rm min}=0.1$\,M$_{\odot}$ and 
M$_{\rm max}=120$\,M$_{\odot}$. The Salpeter IMF is not valid down 
to ${\rm 0.1\,M_{\odot}}$ (e.g. Salpeter 1955, Scalo 1986, Kroupa 2001),
although the IMF shape at ${\rm M_{\rm init}\la\,0.6\,M_{\odot}}$ 
appears to not have large effects on SBF predictions (see Blakeslee 
et al. 1999 for a detailed discussion).

\subsection{Computing Surface Brightness Models}
\label{comp_sbf}

We compute integrated and fluctuation magnitudes and colours, through the 
2MASS near--IR filters, of single--burst stellar populations. In order to 
cover the range of typical ages and metallicities of stellar populations 
in galaxies and globular clusters, we compute the SBF absolute magnitudes 
and colours for ages ranging from 12~Myr to 16~Gyr and initial stellar 
metallicities of ${\rm Z/Z_{\odot}=1/50,1/5,1/2.5,1}$, and 2.5.

The fluctuation luminosity is defined as the ratio of the second moment 
of the luminosity function to its first moment, i.e., the integrated 
luminosity. This can be expressed with the following equation: 

\begin{equation}
\barL \equiv \frac{\Sigma n_i{L_i}^2}{\Sigma n_i L_i}, 
\label{eqsbf}
\end{equation}

\noindent where $n_i$ is the number of stars of type $i$ and luminosity 
$L_i$. Bright stars are the main contributors to the numerator, while 
faint stars contribute significantly to the denominator. The numerator 
of eq.\ \ref{eqsbf} is calculated by summing $L^2$ over all stellar 
types in the models, i.e., $n_i \neq 1$ in general. The denominator 
of eq.\ \ref{eqsbf} and the integrated quantities are obtained also 
by summing the $L$ of all types of star at each wavelength. 

\begin{figure*}
\includegraphics[clip=,width=0.32\textwidth]{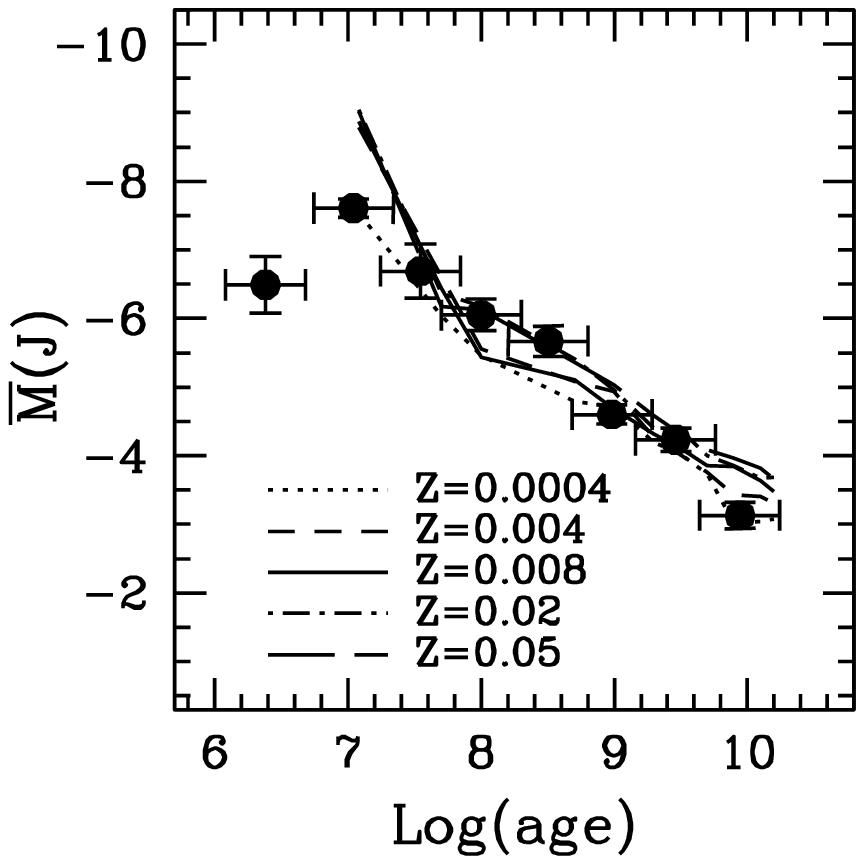}
\includegraphics[clip=,width=0.32\textwidth]{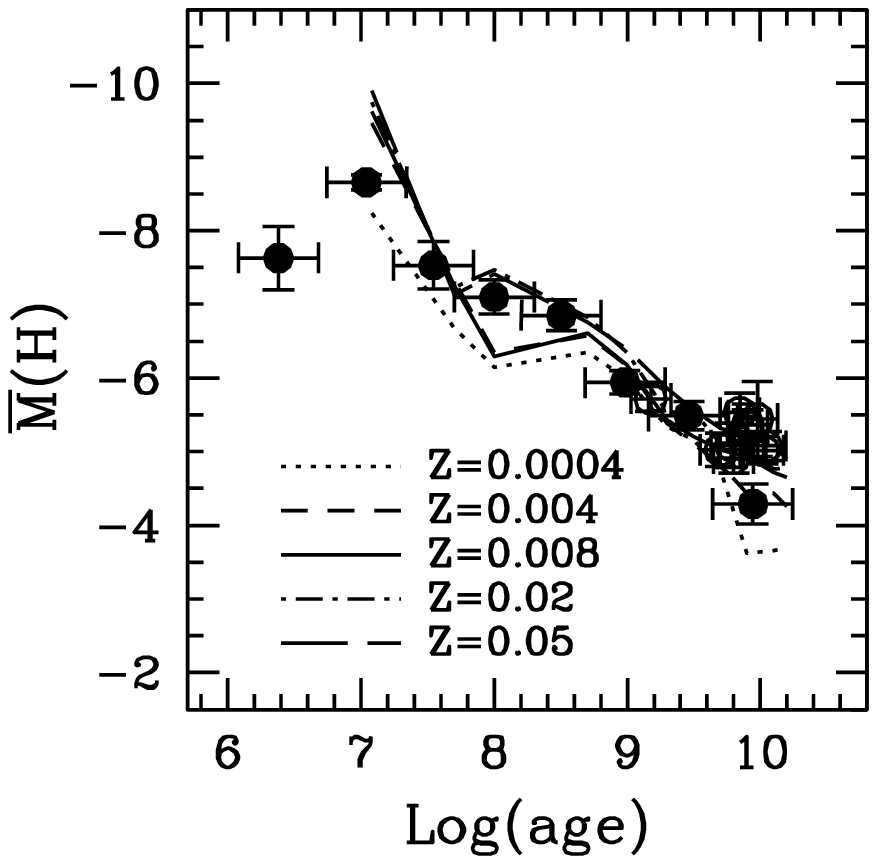}
\includegraphics[clip=,width=0.32\textwidth]{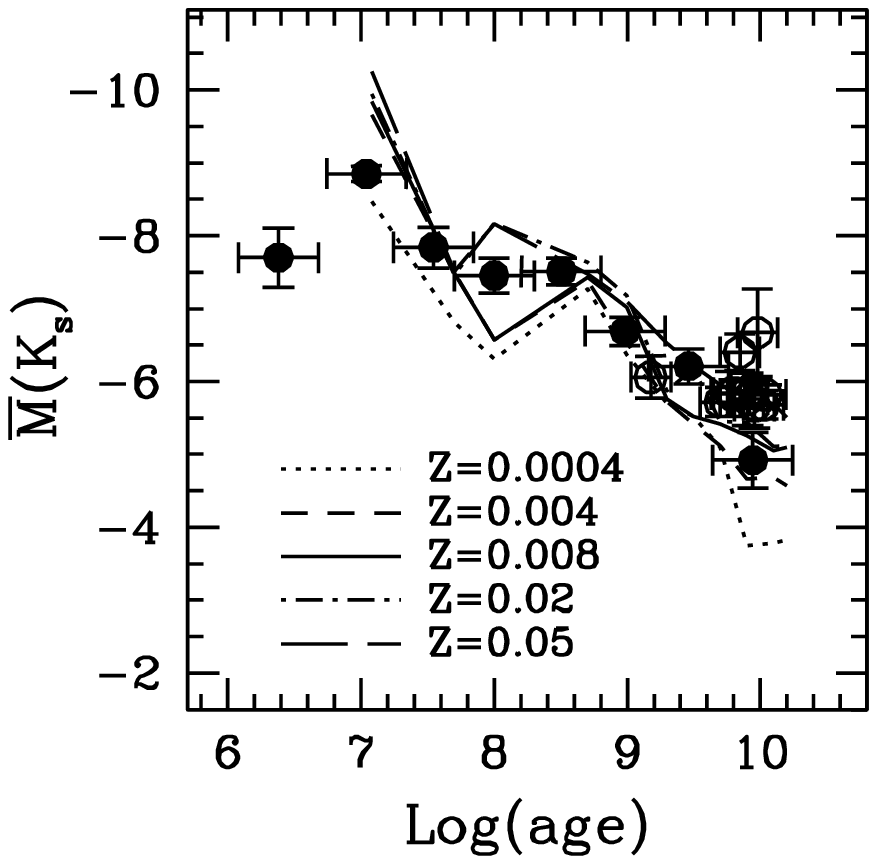}
\caption{Comparison of $J$-band ({\it left}), $H$-band ({\it middle}), 
and $K_s$-band ({\it right}) SBF magnitudes versus age with the 
predictions of stellar population synthesis models. Filled circles 
show the MC supercluster SBF measurements, open circles show those 
derived for Fornax Cluster galaxies.  }
\label{sbf_mag_age}
\end{figure*}

\begin{figure*}
\includegraphics[clip=,width=0.32\textwidth]{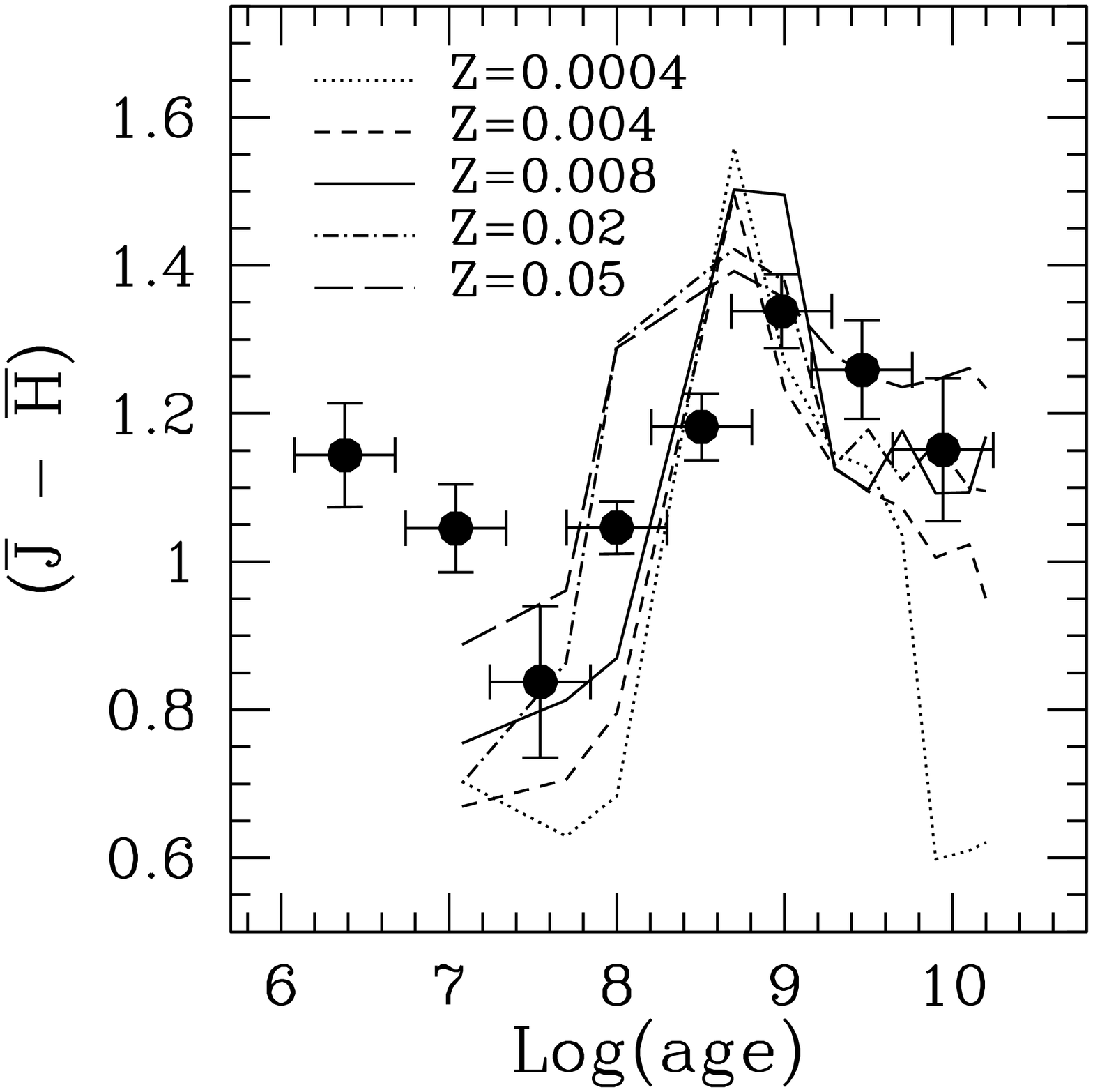}
\includegraphics[clip=,width=0.32\textwidth]{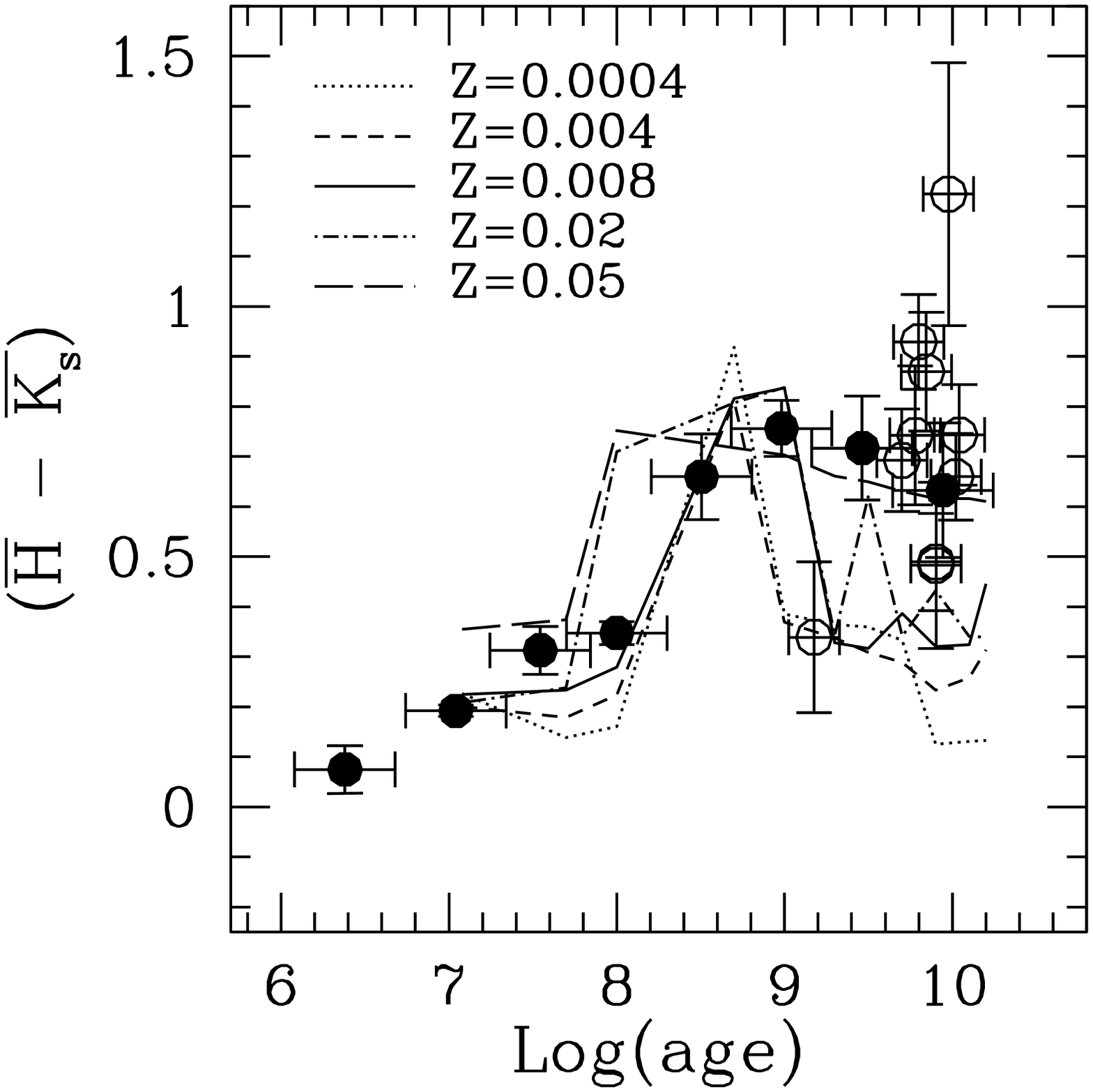}
\includegraphics[clip=,width=0.32\textwidth]{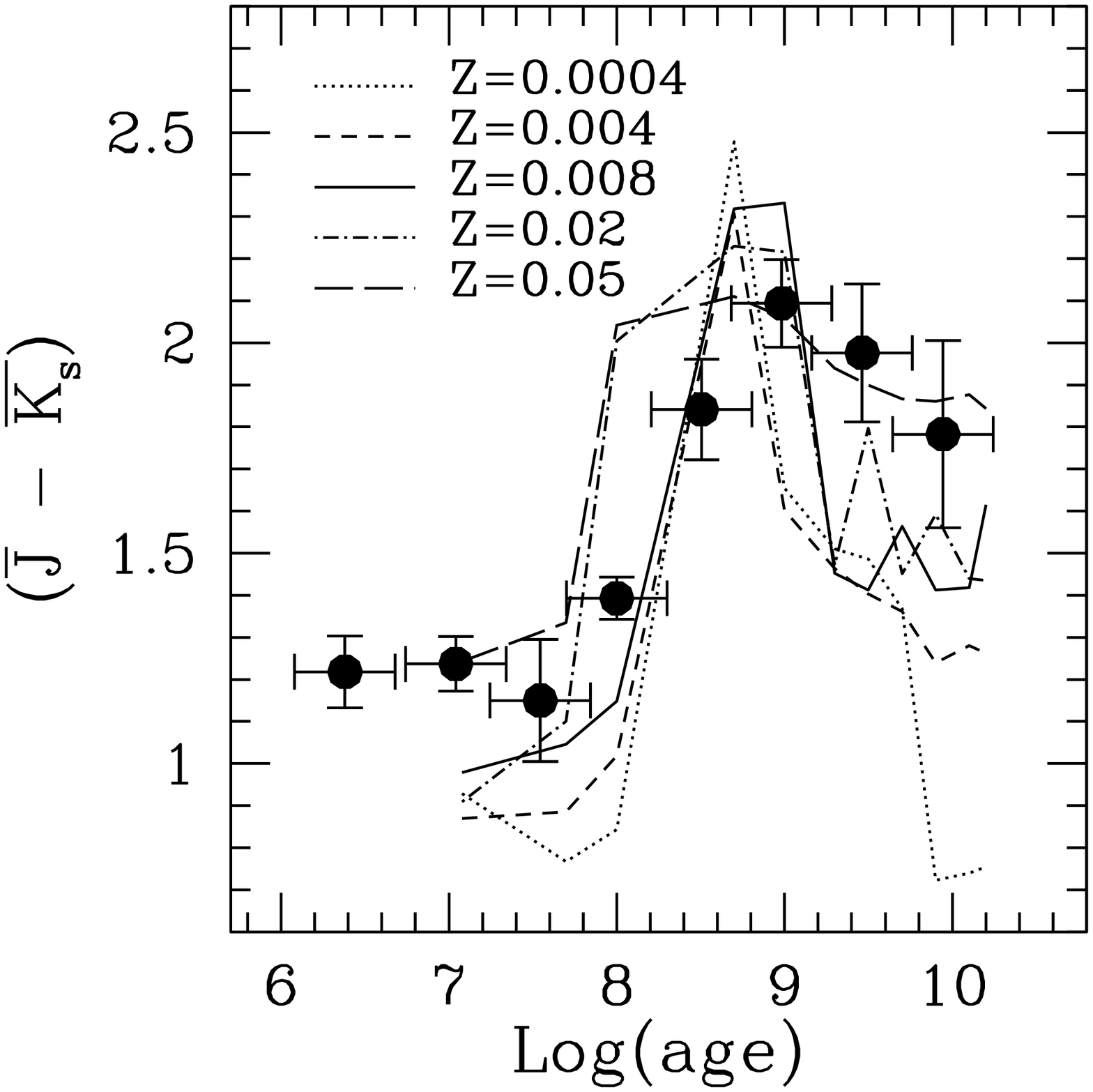}
\caption{Comparison of fluctuation colours $(\barJ-\barH)$ ({\it left}), 
$(\barH-\barKs)$ ({\it middle}), and $(\barJ-\barKs)$ ({\it right}) 
versus age with the predictions of stellar population synthesis models. 
Models and symbols are the same as in Fig.\ \ref{sbf_mag_age}. }
\label{sbf_col_age}
\end{figure*}

\begin{figure*}
\includegraphics[clip=,width=0.32\textwidth]{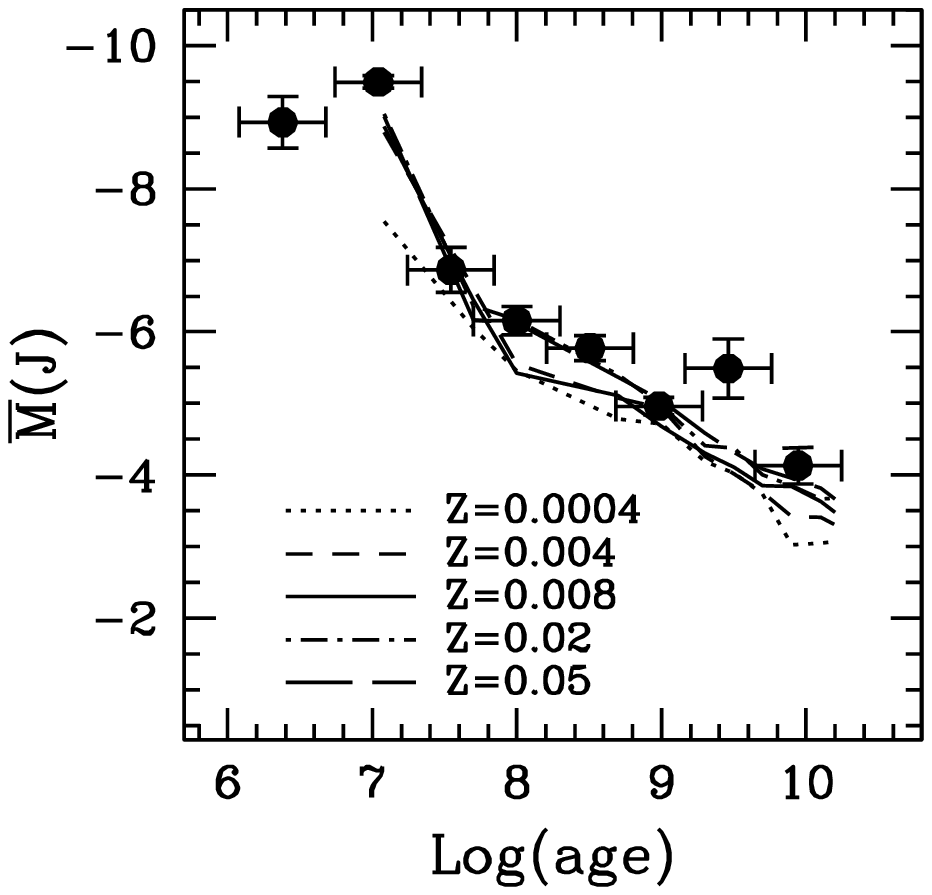}
\includegraphics[clip=,width=0.32\textwidth]{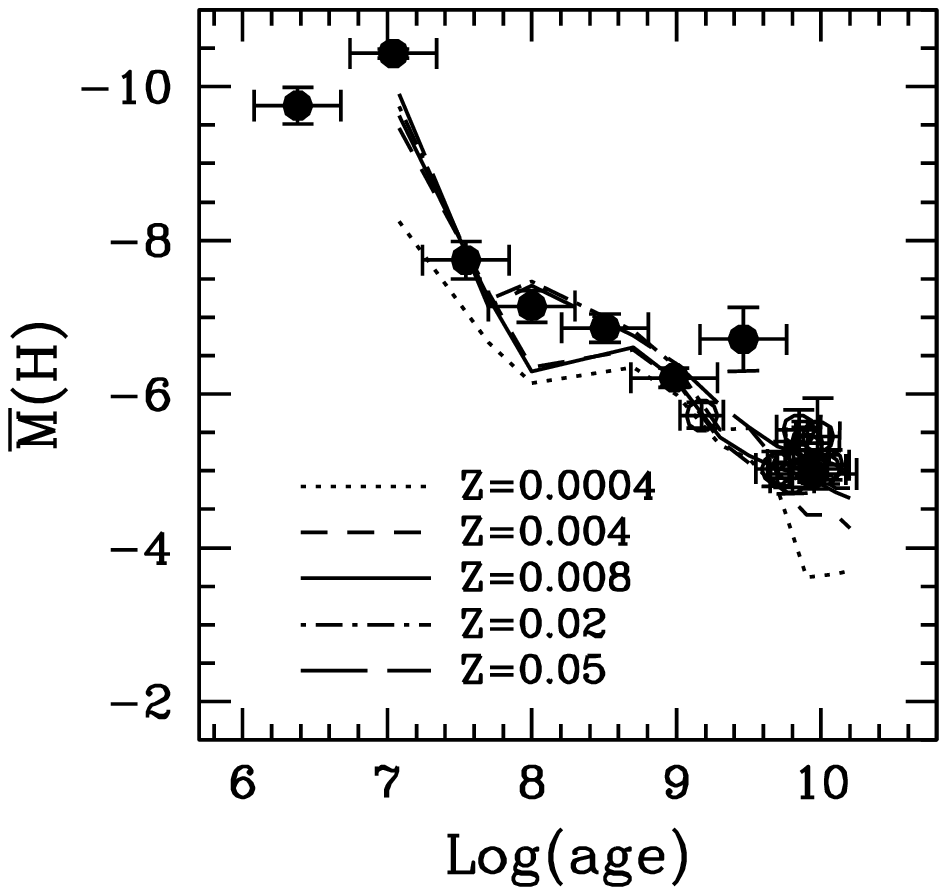}
\includegraphics[clip=,width=0.32\textwidth]{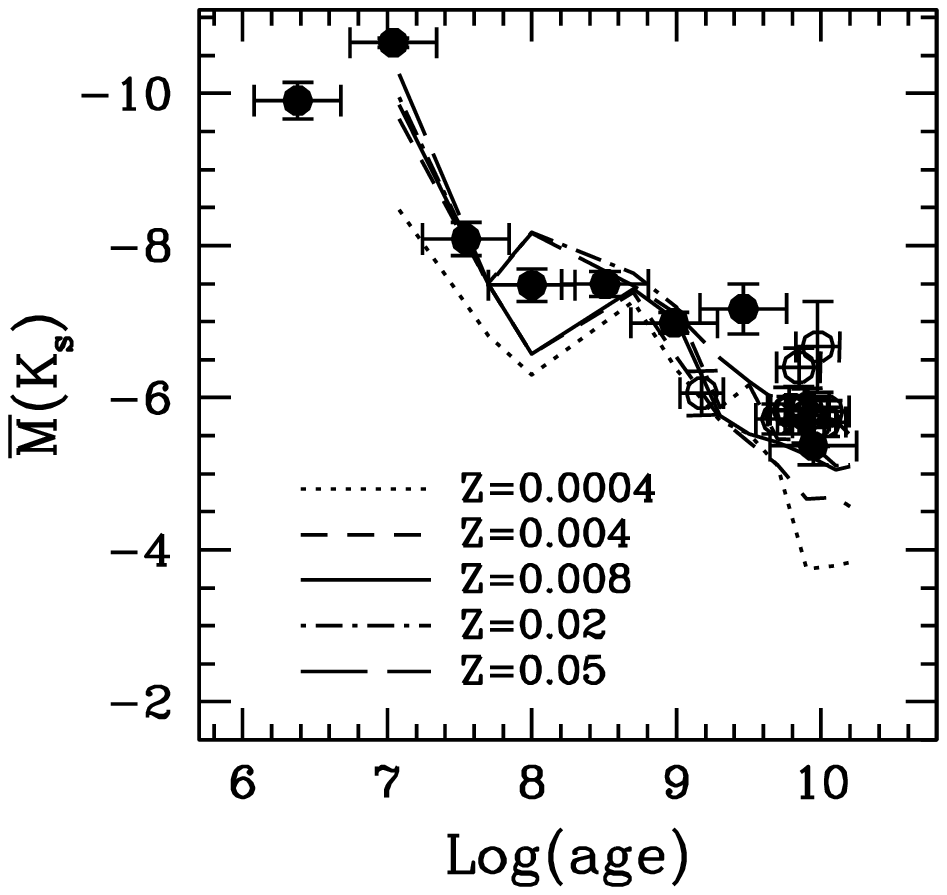}
\caption{Similar to Fig.\,\ref{sbf_mag_age}, but the observed MC 
near-IR SBF magnitudes are estimated including all point sources in 
the 2MASS Point Source Catalog within 1\arcmin\ from the centres of 
the superclusters (see text for more details).  }
\label{sbf_mag_age_all_stars}
\end{figure*}

\begin{figure*}
\includegraphics[clip=,width=0.32\textwidth]{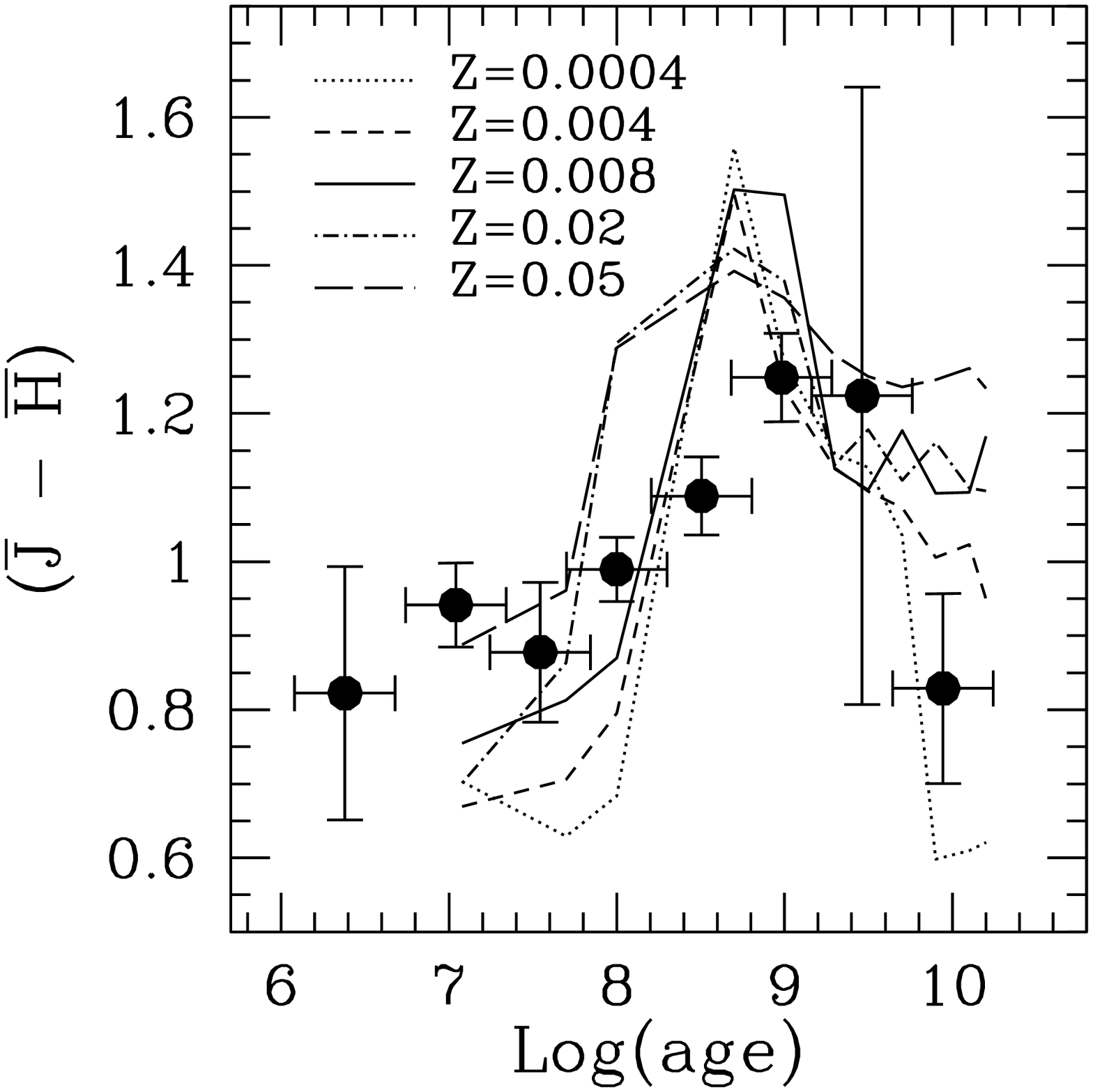}
\includegraphics[clip=,width=0.32\textwidth]{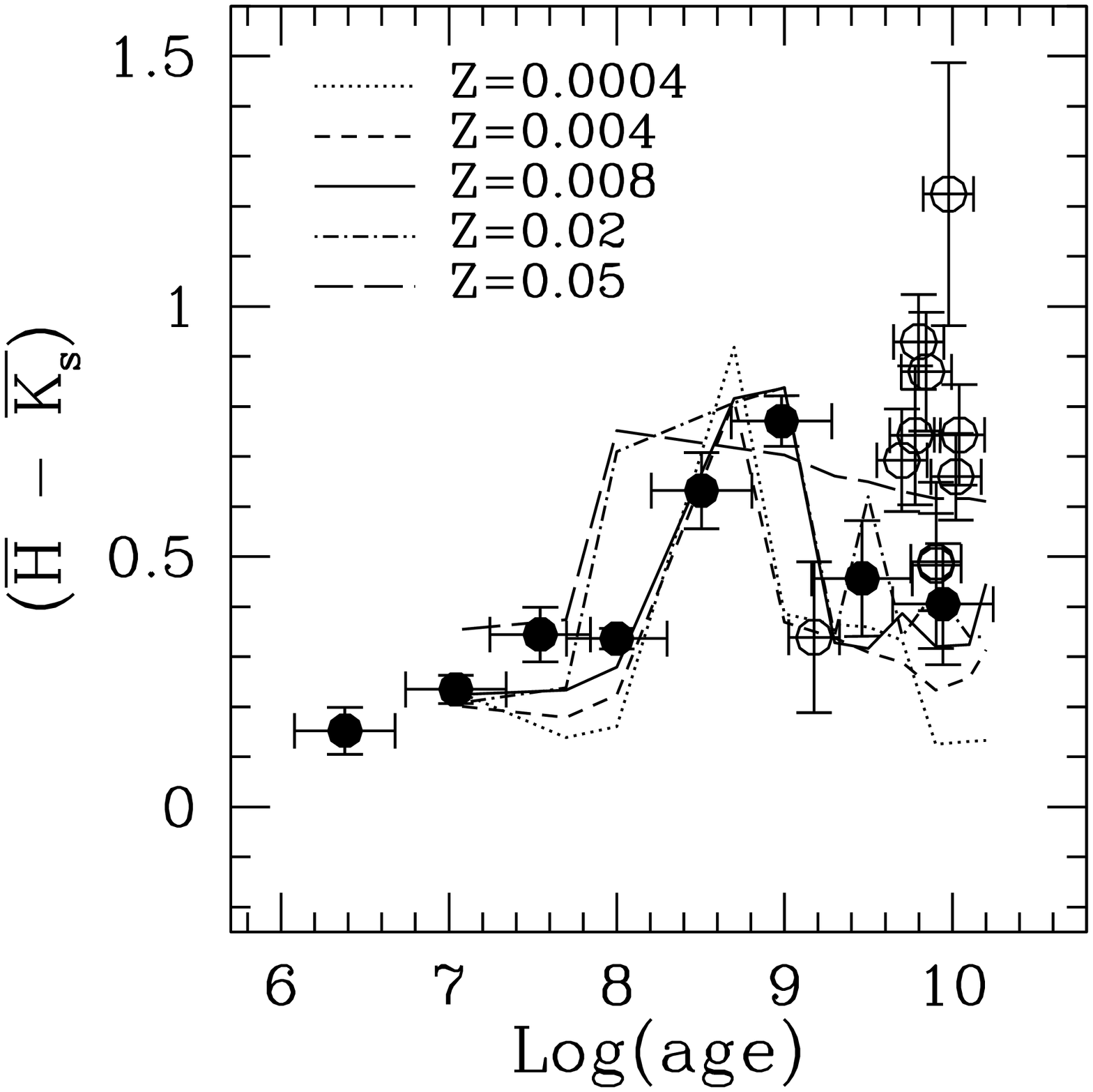}
\includegraphics[clip=,width=0.32\textwidth]{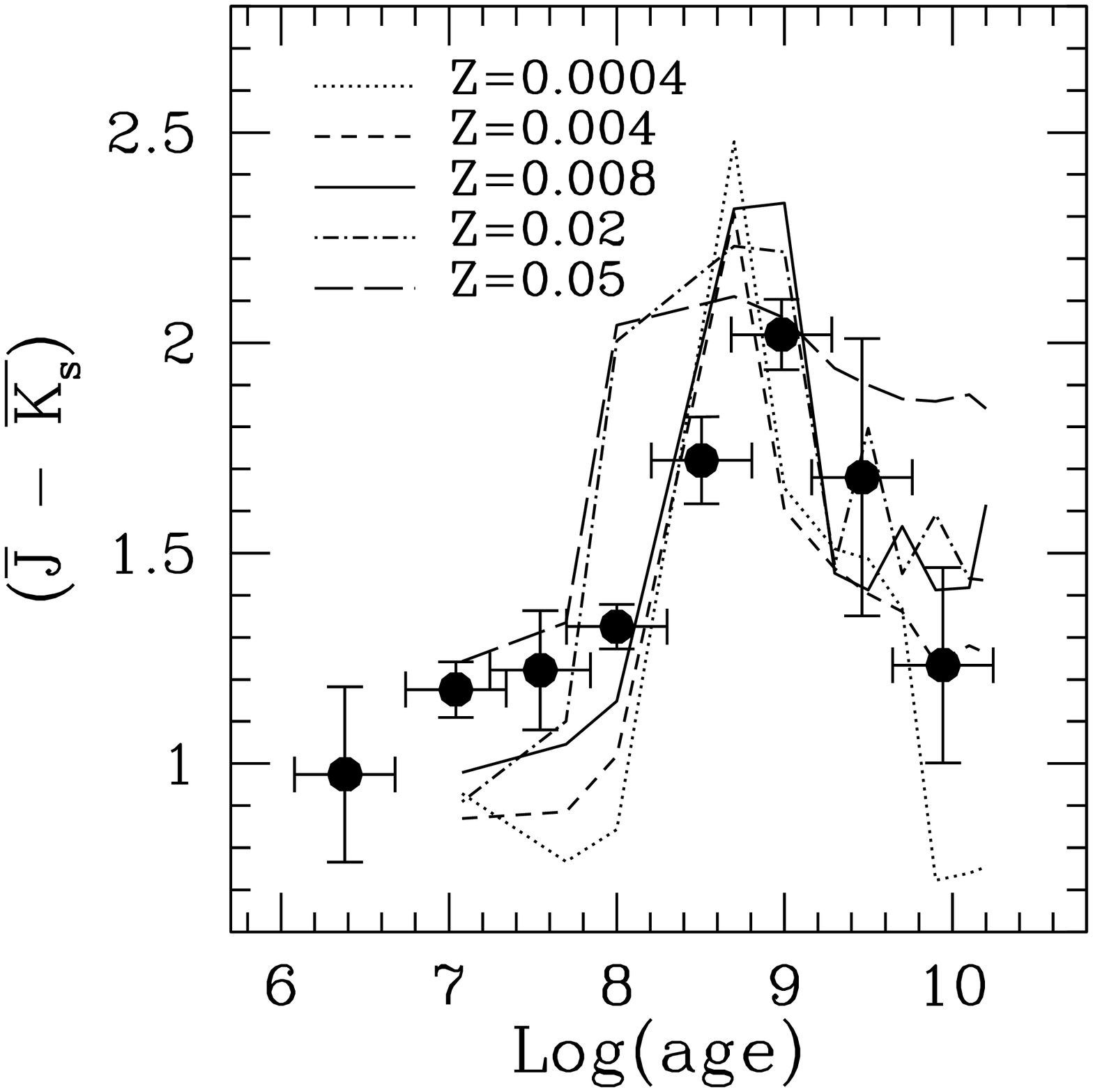}
\caption{Similar to Fig.\,\ref{sbf_col_age}, but the observed MC 
near-IR SBF magnitudes are estimated including all point sources in 
the 2MASS Point Source Catalog within 1\arcmin\ from the centres of 
the superclusters (see text for more details). The match between the 
predictions and the observations for old stellar populations, i.e., 
MCs with SWB type VI and VII, is very good. }
\label{sbf_col_age_all_stars}
\end{figure*}

\begin{figure*}
\includegraphics[clip=,width=0.46\textwidth]{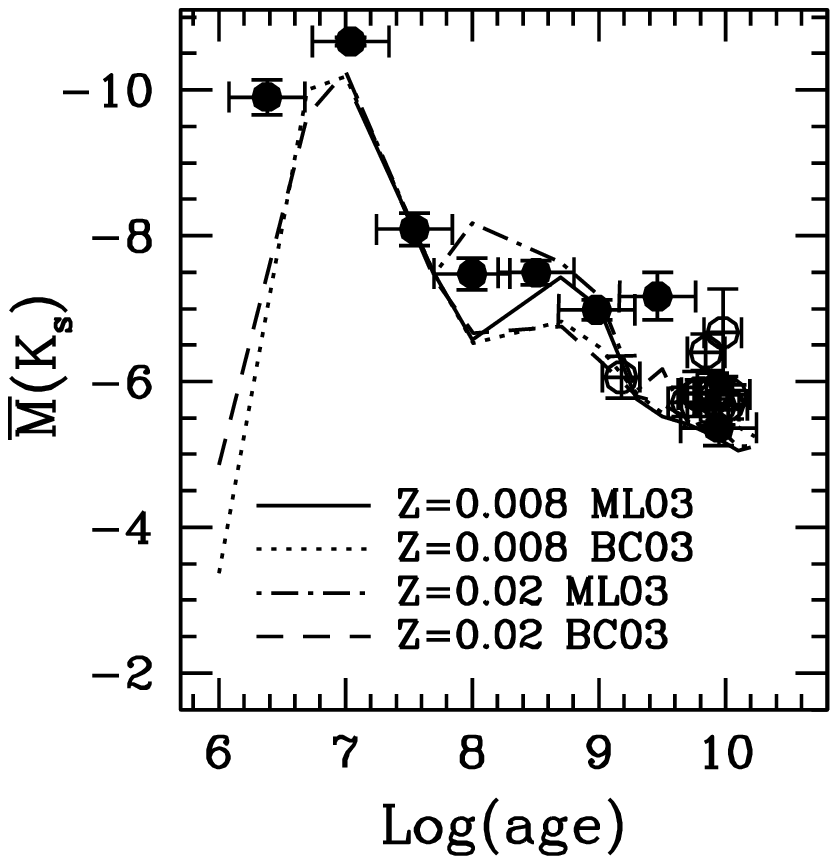}
\includegraphics[clip=,width=0.51\textwidth]{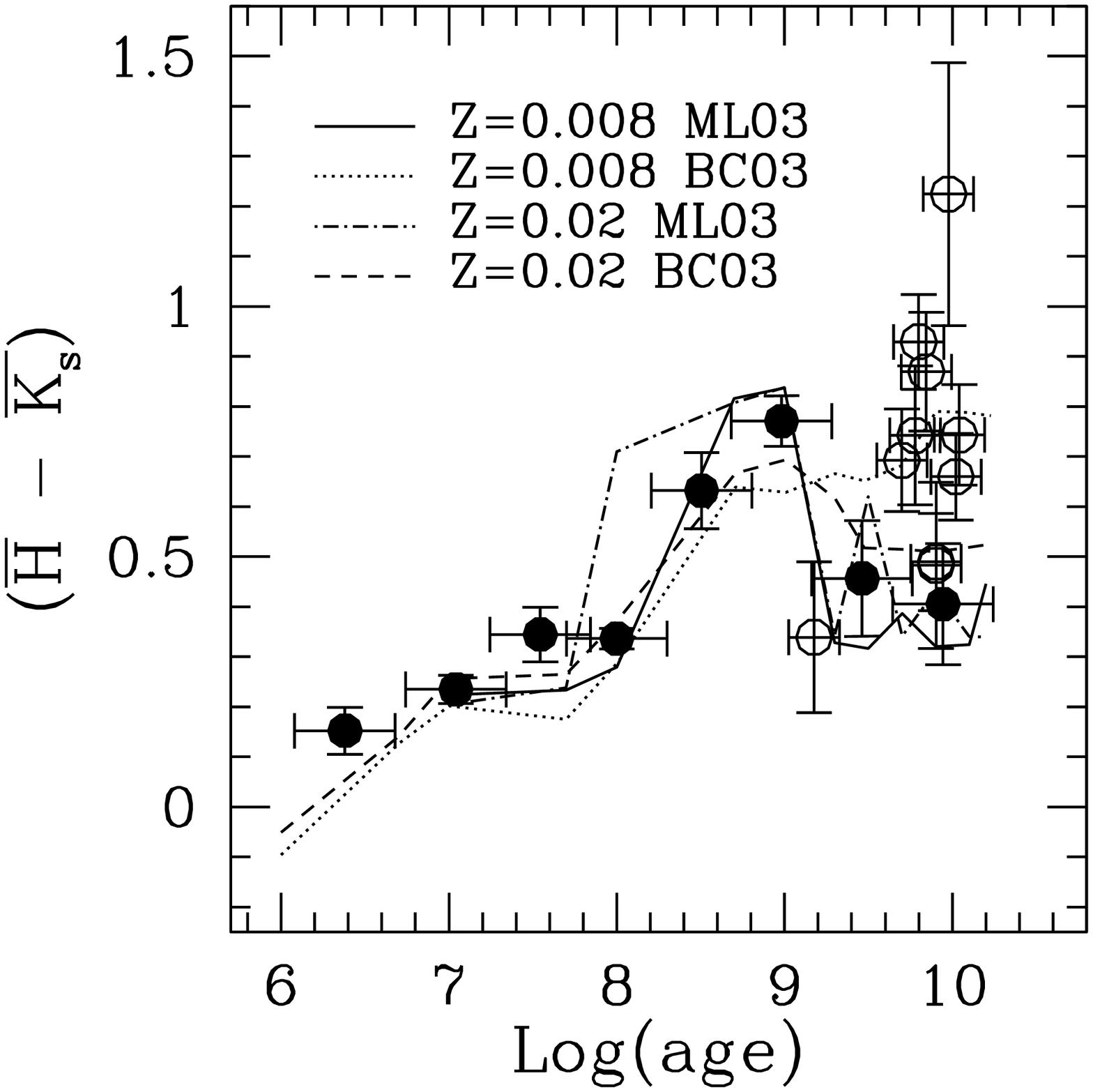}
\caption{Comparison of the temporal evolution of $K_{s}$-band SBF 
magnitude ({\it left}) and fluctuation colour $(\barJ-\barH)$ ({\it 
right}), for both solar and ${\rm Z/Z_{\odot}=1/2.5}$ metallicities.
The observed MC near-IR SBF magnitudes are estimated including all 
point sources in the 2MASS Point source Catalog within 1\arcmin\ 
from the centres of the superclusters (see text for more details). 
{\it Solid and dotted--short dashed lines}: models used in this 
paper, based on isochrones of Mouhcine \& Lan\c{c}on (2003; ML03). 
{\it Dotted and dashed lines}: models used in Gonz\`alez et al (2004),
based on isochrones of Bruzual \& Charlot (2003; BC03). } 
\label{sbf_comp}
\end{figure*}

\section{Results}
\label{comp}

The least ambiguous test for SSP models is to compare there predictions
to the observed properties of globular clusters since they are thought
to be the best observational counterpart of theoretical SSPs. In this 
section, we present a comparison between the predicted and observed 
evolution of near-IR SBFs as a function of age and metallicity.
The corresponding numerical data of the models presented in this paper 
are listed in Table\,\ref{models_tab}.

Gonz\'alez et al. (2004) presented near-IR SBFs for a sample of $191$ 
MC star clusters. In order to reduce stochastic effects due to small 
numbers of stars on fast evolutionary phases, e.g., Red Giant Branch 
(RGB) and AGB, near-IR SBFs for {\it superclusters} were built by coadding 
clusters in the Elson \& Fall (1985, 1988) sample that have the same SWB 
class (Searle, Wilkinson, \& Bagnuolo 1980). Eight superclusters were thus 
assembled, one for each of the seven different SWB classes (Searle et al. 
1980), plus one ``pre-SWB-class'' supercluster. The SWB classification 
is a smooth sequence of increasing age and decreasing metallicity. 
Ages and metallicities of the superclusters were taken from Frogel et al. 
(1990), with the following exceptions: the metallicity of MC superclusters
with SWB type I and II were taken from Cohen (1982); in the case of the 
youngest (pre-SWB) supercluster, its metallicity was extrapolated from 
Fig.\,2 of Cohen (1982), and an age of 2.4~Myr was assumed. 
The MC supercluster ages from Frogel et al. (1990) have been corrected 
to the LMC distance modulus ${\rm (m-M)_{\circ}=18.5}$, used by Gonz\'alez 
et al. (2004). 

Unfortunately, no MC globular cluster with a solar or a super-solar 
metallicity is known. One needs to look at early-type galaxies to test 
the accuracy of SBF predictions for metal-rich stellar populations. 
For this purpose, we use the near-IR SBF measurements for a sample of 
Fornax Cluster early-type galaxies from Liu et al. (2002) and Jensen 
et al. (2003). Ages and metallicities for the sample galaxies have been 
taken from Kuntschner (1998), and are based on the analysis of optical 
spectral features. It is worth to mention that both Fornax Cluster 
galaxy ages and metallicities are luminosity weighted.

\subsection{Fluctuations versus Age}
\label{sbf_age}

\subsubsection{Comparison with observational data}

Fig.\ \ref{sbf_mag_age} shows a comparison between the predicted and 
observed evolution of near-IR absolute fluctuation magnitudes as a 
function of stellar population age. MC globular cluster are shown 
as filled circles,\footnote{SBF magnitudes for the supercluster class 
SWB I are about 0.15 mag brighter, and those for class SWB II about 
0.05 mag fainter than formerly published --Gonz\'alez, Liu, \& Bruzual 
(2005).} and Fornax Cluster galaxies as open circles. 
SSP models with different metallicities are shown with different lines 
as indicated in each panel. The data show that the near-IR absolute 
fluctuation magnitudes get fainter as the superclusters age. The new 
near-IR SBF models agree remarkably well with the observed trend. 
The figure shows clearly that the distance modulus to the MCs derived 
from the predicted near-IR absolute fluctuation magnitudes is in a good 
agreement with the recent determinations from other distance indicators. 
Note that the good agreement between observed and predicted trends is 
relevant considering that the set of free parameters used to describe 
the evolution of TP-AGB stars has not been tuned to reproduce the 
observed fluctuation magnitudes. This confirms the reliability of the 
adopted theoretical tool to model the evolution of TP-AGB stars, and 
then the stellar population models used to predict near-IR SBFs. 

For stellar populations older than 100~Myr, the evolution of low- 
and intermediate-mass stars through the red giant branch and the AGB 
regulates the evolution of near-IR spectrophotometric properties and 
SBF signal. The monotonic dimming of near-IR fluctuation magnitudes 
in these stellar populations is related to the evolution of late-type 
giant star content. As a stellar population ages, the mass of stars 
fueling the evolution of near-IR SBF signal, i.e., AGB stars for ages 
younger than $\sim\,1.5\,$Gyr, and red giant stars for older ages, 
decreases. At a fixed metallicity, the average luminosity of stars 
on the red giant branch decreases with a stellar population age (see 
Gonz\'alez et al.\ [2004] for a detailed discussion). 
On the other hand, SSP models show, at a fixed age, brighter absolute 
near-IR fluctuation magnitudes at higher stellar population metallicity.
Due to opacity effects, more metal-rich stars are redder than their 
metal-poor counterparts and, hence, are brighter in the near-IR.

On top of the overall trend of near-IR SBFs dimming as stellar 
populations age, a short migration of near-IR fluctuations to brighter 
magnitudes is predicted for stellar populations between 0.5 and 
$\sim\,1.5$~Gyr. 
The predicted migration gets brighter going from $J-$band to $K-$band. 
This behaviour is due to a temporary predominance of bright and cool 
carbon stars. As the production rate of such stars decreases 
significantly at high metallicity, the migration is virtually absent 
for super-solar stellar population models. 

For stellar population studies, although the fluctuation absolute 
magnitudes can be used to set constraints on the properties of the 
stellar populations, the fluctuation colours are, in principle, more 
useful, as they do not depend on the distance determinations. It is 
therefore interesting to investigate their evolution and sensitivity 
to stellar population properties. In Fig.\ \ref{sbf_col_age} we show 
a comparison between the observed and predicted temporal evolution 
of near-IR fluctuation colours. Different symbols and lines are the 
same as in Fig.\ \ref{sbf_mag_age}. 

The new SSP models show that near-IR fluctuation colours get redder 
until they reach a maximum at $0.7-1.0$~Gyr, depending slightly on 
metallicity. Subsequently, the evolution of near-IR fluctuation 
colours reverses and they get bluer up to an age of $1.5-2.0$~Gyr. 
From then on, except for the most metal-poor models (${\rm 
Z/Z_{\odot}=1/50}$), near-IR fluctuation colours of older stellar 
populations are almost constant over several Gyrs. This is not 
completely unexpected. The predicted evolution is quantitatively 
similar to what is anticipated for near-IR integrated colours 
(Mouhcine \& Lan\c{c}on 2003). The only difference 
is that, for sub-solar SSP models and for a given age, integrated 
colours get redder for higher metallicities, while the migration 
of near-IR fluctuation colours to the red is sharper for metal-poor 
stellar populations. The most metal-poor SSP model shows the reddest 
near-IR fluctuation colours at the end of this evolutionary phase. 
The predicted sensitivity of near-IR fluctuation colours to 
metallicity is due to (i) a larger sensitivity of near-IR SBF 
signal to the presence of bright and cool AGB stars, and (ii) a 
higher carbon stars formation efficiency at low metallicity. 
The match between the predicted near-IR fluctuation colours and 
the observed ones for MC superclusters of SWB types lower than V 
is good. The observed abrupt jump of near-IR fluctuation colours 
to the red for stellar populations between 100~Myr and 1~Gyr is 
well matched by the predicted evolution.  

The observed near-IR fluctuation colours for MC superclusters of 
SWB types VI and VII are surprisingly redder than the predicted 
fluctuation colours given their metallicities, i.e., ${\rm 
0.008\la\,Z\,\la\,0.0006}$. The near-IR light budget of stellar 
populations older than $\sim\,3$ Gyr, i.e., globular clusters of 
SWB type VI or larger, is dominated by red giant stars; the physics 
of these stars is well understood to the level of detail that 
matters for stellar population synthesis models, and its effects 
on integrated properties of stellar populations are well understood.
Changing the SWB class-age transformation does not solve the 
discrepancy between the predicted (blue) and observed (red) near-IR 
fluctuation colours.
It is worth to mention that for both superclusters, a good agreement 
is found between observed and predicted integrated colours given 
their ages and metallicities. It is hard to come up with any physical 
explanation for the observed red fluctuation colours of the two 
oldest MC star clusters. One possible solution to the discrepancy 
between the observed red and the predicted blue fluctuations for 
metal-poor and old stellar populations is that the measured near-IR 
SBFs of the two oldest MC superclusters, i.e., SWB types VI and VII, 
are not well determined for observational reasons. The bright end 
of the RGB sequences of MC superclusters of SWB types VI and VII 
is poorly populated (see figure $3$ of Gonz\'alez et al. 2004). 
A possible contamination of the upper bright end of the red giant 
branch by foreground red stars might affect strongly the measured 
near-IR SBF magnitudes given their sensitivity to the presence of 
red stars. 

A second potential source of uncertainty is the methodology used by
Gonz\'alez et al. (2004) to estimate near-IR SBF magnitudes for MC 
star clusters. They have eliminated all point sources with dubious 
photometry using the flags from the 2MASS Point Source Catalog.
%
Table\,\ref{sbf_tab} lists the near-IR fluctuation absolute 
magnitudes for the MC superclusters when all point sources 
within a radius of 1\arcmin\ from the centres of the superclusters 
are included, independently of the 2MASS Point Source Catalog 
flags. A comparison with the near-IR SBF absolute magnitudes 
listed in Table\,$4$ of Gonz\'alez et al. (2004) and revised 
in Table\,$1$ of Gonz\'alez et al. (2005) shows that the near-IR 
SBF absolute magnitudes of some of the MC superclusters, i.e., 
Pre-SWB, I, VI, and (albeit to a lesser extent and mostly in the
$J$-band) VII, are significantly affected by the procedure used 
to select the members of the superclusters, and by their 
photometric quality. Fig.\,\ref{sbf_mag_age_all_stars} shows a 
comparison between the predicted evolution of near-IR fluctuation 
absolute magnitudes, and observed ones for the MC superclusters 
as listed in Table\,\ref{sbf_tab}.

The systematic errors incurred when deriving SBF magnitudes of 
resolved star clusters have been amply discussed in Gonz\'alez 
et al. (2004; see their Sect. 4.2, and Fig.\ 5). The two main 
issues involved are the sky level and, notoriously, the crowding. 
Through the blending of sources, crowding will make SBF magnitudes 
brighter. Crowding is most likely in small areas with relatively 
bright stars. Consequently, the central regions of star clusters 
are always the most vulnerable to crowding and, if the problem 
indeed exists, the fluctuation magnitudes should in principle be 
brighter there than when including larger radii. 

The analysis by Gonz\'alez et al. (2004),\footnote{The analysis
consisted in the comparison of fluctuation magnitudes and colours
calculated in different regions of the MC superclusters and, 
especially, between values obtained in the central 0\arcsec\ to
20\arcsec\ and from the whole region encompassed between 0\arcsec\
and 60\arcsec.} updated to take into account the corrected data
of SWB classes I and II (Gonz\'alez et al.\ 2005), indicates that 
in all but the Pre-SWB and type I superclusters, when using only
stars with good photometry, fluctuation magnitudes in the central 
regions are actually fainter, not brighter, than when considering 
whole regions with 0\arcsec\ $\le r \le$ 60\arcsec. That is, while 
the Pre-SWB and type I superclusters show the expected bias, the 
other ones display the effects of crowding in their centers only 
indirectly, since blended sources are discarded via the photometric 
flags and hence SBFs appear fainter. When carrying out the analysis 
with all the stars, the fluctuation magnitudes for the whole regions 
between 0\arcsec\ and 60\arcsec\ change significantly only for the 
superclusters types Pre-SWB, I, and VI. 
And, tellingly, only for those three superclusters the fluctuations 
for the centralmost regions are now (when all sources are considered) 
brighter than for the whole regions between 0\arcsec\ and 60\arcsec. 
The implication is that the Pre-SWB and type I superclusters suffer 
from a crowding so severe that it leaves its expected, direct 
imprint even when taking into account only sources with nominally 
good photometric measurements. On the other hand, for supercluster 
type VI, crowding is only a problem when all point sources are 
included. Ideally, one should derive fluctuation magnitudes from 
uncrowded regions, but in the case of these data the 2MASS flags 
have been useful to correct the effects of mild crowding. 
Unfortunately, we have to conclude that the measurements for the 
superclusters Pre-SWB and type I have to be taken with caution 
until data with better spatial resolution can be obtained. 
In the case of the supercluster type VII, even when it does not
appear to be badly crowded, the brightest end of old superclusters
is poorly populated, so adding or substracting a few bright stars
could have disproportionately large effects for statistical 
reasons. Aside from crowding, this ingredient could also be at 
play in supercluster type VI.
  
Interestingly, fluctuation colours do not seem to be affected by 
crowding. Fig.\,\ref{sbf_col_age_all_stars} shows a comparison 
between the predicted evolution of near-IR fluctuation colours, 
and those estimated for the MC superclusters using the near-IR 
SBF absolute magnitudes listed in Table\,\ref{sbf_tab}. 
For regions within 1\arcmin\ of their centres and for all 
superclusters, the empirical values derived from the observations 
including all sources are consistent within 1 to 2--$\sigma$ errors 
with those obtained when only sources with good photometry are 
taken into account. To explain the simultaneous observations 
that fluctuation magnitudes are sensitive to crowding, while 
fluctuation colours do not seem to be, it is possible to suppose 
that if crowding is similar at two wavelengths, its effect on the 
fluctuation magnitudes will cancel out when calculating a colour. 
Alternatively, the colour of two blended stars could be dominated 
by the bright star, and yet the fainter star would contribute 
significantly to the measured brightness. This would be similar 
to an effect that has a role in the skewed shape of the luminosity 
distribution of main sequence stars with a given spectral type 
(see, e.g., Binney \& Merrifield 1998).
In that case, the departure from the Gaussian shape is produced
at least in part by unresolved binaries in the sample. The spectral
classification is likely to be based on the brighter star, while
both stars will contribute to the measured brightness. In the case
of fluctuation measurements, blended stars are not real binaries,
but their colour (like the spectral class of main sequence binaries) 
might be dominated by the bright star, while the fainter star will 
still contribute significantly to the detected flux.

Finally, Fig.\,\ref{sbf_col_age_all_stars} suggests that there 
might be a better match between predicted and observed near-IR
colours when all sources, regardless of their photometric quality,
are included. We caution, however, against giving too much weight
to this result. For example, although the very crowded Pre-SWB 
and SWB type I superclusters, as well as the mildy crowded type 
VI supercluster, move in the {\it right} direction, however 
within the errors with respect to the values estimated without 
point sources with faulty photometry, for $(\barJ-\barH)$ and 
$(\barJ-\barKs)$ the overall impression of improvement hinges 
mostly on the point for SWB VII supercluster. The discussion 
in this section points towards the conclusion that the estimated 
near-IR SBF magnitudes of the two youngest and the two oldest 
MC superclusters by Gonz\'alez et al. (2004) might be affected 
by important observational uncertainties. One could perform 
simulations to test our hypotheses on the resilience of
fluctuation colours to crowding but, mainly, it would be definitely 
worthwhile to reobserve very young and very old MC star clusters 
in the near-IR with better spatial resolution and sky stability, 
in order to determine properly their near-IR SBF magnitudes.
%
%

The agreement between observed near-IR fluctuation colours of Fornax 
Cluster galaxies and predicted ones for similar ages and metallicities 
is satisfactory, given the SBF measurement errors. However, a few 
early-type galaxies show much redder colours than what is predicted 
by the most metal-rich SSP model. Note that observed near-IR 
integrated colours of Fornax Cluster galaxies agree well with the 
predictions of SSP models given their ages and metallicities. 
The star formation histories of Fornax Cluster galaxies are likely 
different from the single burst history assumed in the models, and 
intermediate age stellar populations can exist (e.g., Bower et al. 
1998). 
Various evidence suggests that early-type galaxies have had more 
complex star formation histories, and may contain a mixture of 
stellar populations with different ages and metallicities. 
The modelling of the effects of complex star formation history on 
the evolution of near-IR fluctuation colours is beyond the scope of 
this paper. Liu et al. (2002) have shown that recent star formation 
events on the top of an old stellar population are suspected to 
produce a scatter of near-IR fluctuations properties. An interesting 
prediction is that after few Gyr, when the effect of a recent star 
formation event on integrated colours almost vanishes and these become 
typical of old stellar populations, a sizeable signature of burst 
remains on the near-IR fluctuation properties. If a fraction of the 
galaxy stellar mass was formed in the last few Gyr, it can produce a 
significant spread of near-IR SBFs, depending on the mass fraction 
involved in the burst and its metallicity. This might explain why 
observed integrated colours of Fornax cluster galaxies are well 
reproduced while some Fornax cluster galaxies show redder near-IR 
fluctuation colours than the reddest models. The observed discrepancy 
might indicate that the single burst model is inappropriate to model 
their star formation histories. 


\subsubsection{Comparison with theoretical predictions}

Gonz\'alez et al. (2004) have calculated near-IR SBF magnitudes and 
fluctuation colours for single burst stellar populations using Bruzual 
\& Charlot (2003) isochrones over similar ranges of age and metallicity
than what is presented in this paper. Fig.\,\ref{sbf_comp} displays a
comparison between the predictions of the temporal evolution of $K_s$-band 
SBF magnitude and $(\barH-\barKs)$ fluctuation colour based respectively 
on Bruzual \& Charlot (2003) and Mouhcine \& Lan\c{c}on (2003) isochrones. 
Also plotted are the observed near-IR SBF magnitudes and fluctuation 
colours of MC superclusters estimated using all point sources detected 
within 1\arcmin\ from the centres of the superclusters. 
Both sets of theoretical predictions are shown for Z=0.02, and Z=0.008 
metallicities. Other sets of theoretical predictions of near-IR SBF 
magnitudes of simple stellar populations are available in the literature; 
however, they generally use Bruzual \& Charlot (2003) isochrones or 
one of their earlier versions (Liu et al. 2001, Mei et al. 2001), or they
are computed for old stellar populations only, i.e., they do not cover 
the stellar population age range spanned by the MC superclusters (e.g., 
Worthey 1993, Blakeslee et al. 2001, Cantiello et al. 2003).

The figure shows that both sets of models predict qualitatively similar 
evolution of $K_s$-band SBF magnitude over the age range comprised by 
them, i.e., the $K_s$-band SBF magnitude gets fainter as a stellar 
population ages. However, sizeable differences between the two sets of 
predictions are apparent. For stellar populations 
between 100~Myr and 1.5~Gyr, the models based on Mouhcine \& Lan\c{c}on 
(2003) isochrones predict a jump of the $K_s$-band SBFs to brighter 
magnitudes, and that this jump is metallicity dependent. Conversely, 
the models based on Bruzual \& Charlot (2003) isochrones show a 
weak brightening of the $K_s$-band SBF magnitude. The predicted 
$K_s$-band SBF magnitudes by Gonz\'alez et al. (2004) at the age of MC
superclusters of SWB types III and IV, for which the SBF measurements 
are not affected seriously by observational uncertainties, are fainter 
than the observed ones by $0.5-1$ magnitude. The presence of AGB stars 
in the stellar populations during this age interval does not translate 
into noticeable changes in the near-IR light in these models. 
To reproduce the $K_s$-band SBF magnitudes observed for the MC 
superclusters within this age range using Gonz\'alez et al. (2004) 
models, a super-solar metallicity is needed, i.e., ${\rm 
Z/Z_{\odot} = 2.5}$, much larger than the mean metallicity of MC star 
clusters of these ages (${\rm Z/Z_{\odot}\approx\,1/2}$). To model the 
evolution of intermediate mass stars through the TP-AGB phase, Bruzual 
\& Charlot (2003) have used the TP-AGB lifetimes published by Vassiliadis 
\& Wood (1993). The TP-AGB lifetimes used in Mouhcine \& Lan\c{c}on (2003) 
stellar population synthesis models are larger than those of Vassiliadis 
\& Wood (1993). The total number of TP-AGB stars present in a stellar 
population at a given age is directly proportional to the TP-AGB lifetime 
of the turn-off star at this age (e.g., Renzini \& Buzzoni 1986; 
Mouhcine \& Lan\c{c}on 2002). So, shorter lifetimes underestimate the 
light budget of the TP-AGB star population, and lead to predict faint 
near-IR SBF magnitudes. 

The comparison of the predicted temporal evolution of $(\barH-\barKs)$ 
fluctuation colour by both sets of stellar population synthesis models, 
shown in the right panel of Fig.\,\ref{sbf_comp}, again display 
noticeable differences. The most striking one is that, while models with
Z=0.02 and Z=0.008 based on Mouhcine \& Lan\c{c}on (2003) isochrones 
follow a similar qualitative evolution (i.e., they redden with age 
up to $\sim\,1\,$Gyr, get bluer later on up to $\sim\,2\,$Gyr, and keep 
constant fluctuation colours for older stellar populations), the models 
based on Bruzual \& Charlot (2003) isochrones differ between them. 
For this set, the solar metallicity model shows a qualitative bahaviour 
similar to what of our models; however, their Z=0.008 model predicts 
that the $(\barH-\barKs)$ fluctuation colour gets redder as the stellar 
population ages. The result is that, for older stellar populations, the
predicted $(\barH-\barKs)$ fluctuation colours are redder for Z=0.008 
than for solar metallicity. Note that the predicted $(\barH-\barKs)$ 
fluctuation colour by Gonz\'alez et al. (2004) for old stellar 
populations with a super-solar metallicity, i.e., Z=0.05, are also bluer 
than these of Z=0.008 model. This is the opposite of what is anticipated 
using Mouhcine \& Lan\c{c}on (2003) isochrones. For these models, the 
predicted $(\barH-\barKs)$ fluctuation colours of old stellar populations 
with Z=0.008 are as red as what is observed for Fornax Cluster galaxies, 
for which the mean metallicity is super-solar on average. 
The scaling of $(\barH-\barKs)$ with metallicity for old stellar 
populations predicted by Gonz\'alez et al (2004) cannot be explained 
by the presence of carbon stars, for which the formation efficiency is 
higer at low metallicity. The near-IR properties of stellar populations 
older than a few Gyrs are dominated by RGB stars. These stars show redder 
near-IR colours as their initial metallicities increase. 
Both the isochrones used by Gonz\'alez et al. (2004) and the Mouhcine
\& Lan\c{c}on (2003) isochrones are constructed using the same set of 
stellar tracks, for evolutionary stages earlier than the TP-AGB phase 
for intermediate and low mass stars. The only remaining potential source 
of the discrepancies observed between their predicted fluctuation colours 
is the use of different stellar libraries to transform the isochrones 
from theoretical diagrams to observed ones. Mouhcine \& Lan\c{c}on (2003) 
used the Lejeune et al. (1997, 1998) stellar spectral library, while 
Bruzual \& Charlot (2003) used the library of Westera et al (2002).  

The larger amplitude of the jump toward the red of $(\barH-\barKs)$ 
for stellar populations younger than $\sim\,1\,$Gyr 
predicted by models based on Mouhcine \& Lan\c{c}on (2003) isochrones 
is due to the larger TP-AGB evolutionary phase lifetimes used. 
The jump of near-IR fluctuation colours to the red for stellar 
populations between $\sim\,0.1$ Gyr and $\sim\,1.5$ Gyr, i.e., when the 
near-IR light is dominated by TP-AGB stars, as predicted by Gonz\'alez 
et al. (2004) does not show a large sensitivity to stellar population 
metallicity, even though metal-rich stellar populations do show redder 
fluctuation colours than metal-poor ones (see figure 9 of 
Gonz\'alez et al. 2004). This contrasts with the predictions presented 
in this paper, where the jump to the red of near-IR fluctuation colours 
is sharper for metal-poor stellar populations as discussed above. 
The disagreement is due to the fact that Bruzual \& Charlot (2003) do 
not take into account the sensitivity of carbon star formation rate to 
metallicity for metal-poor stars.


\subsection{Fluctuations versus Metallicity}
\label{sbf_z}

In this section, we compare the predicted and observed relationships 
between fluctuation magnitudes and colours, and stellar population 
metallicity.

\begin{figure*}
\includegraphics[clip=,width=0.32\textwidth]{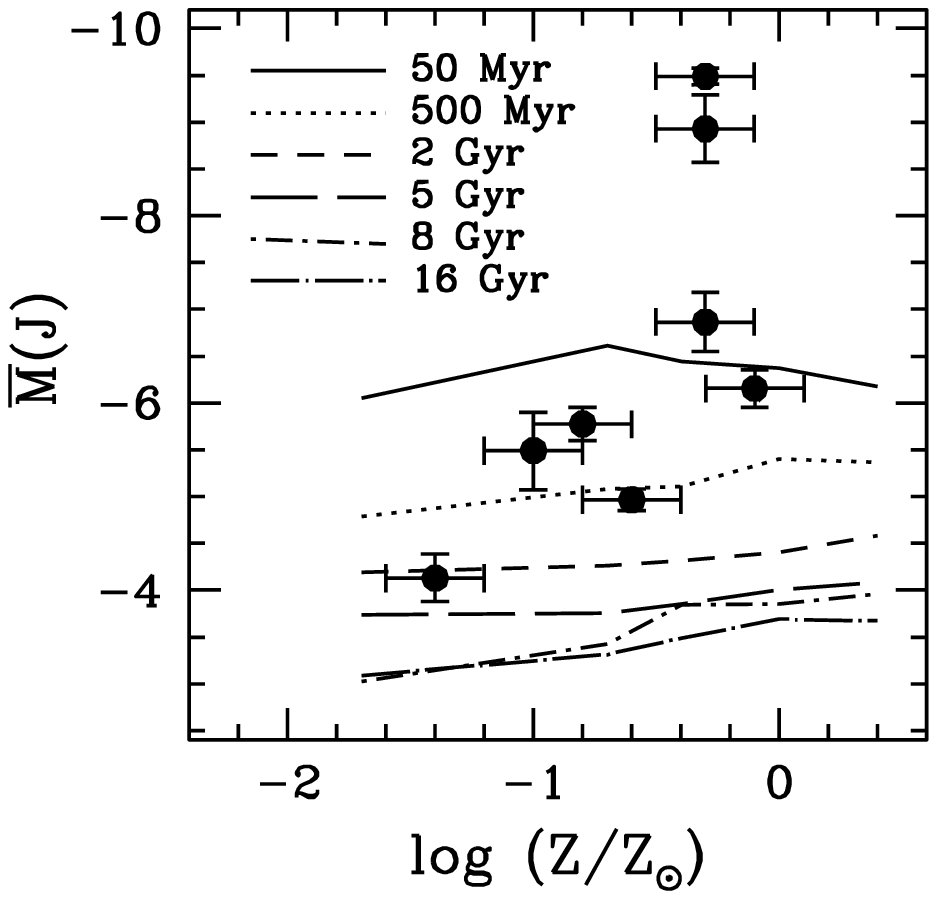}
\includegraphics[clip=,width=0.32\textwidth]{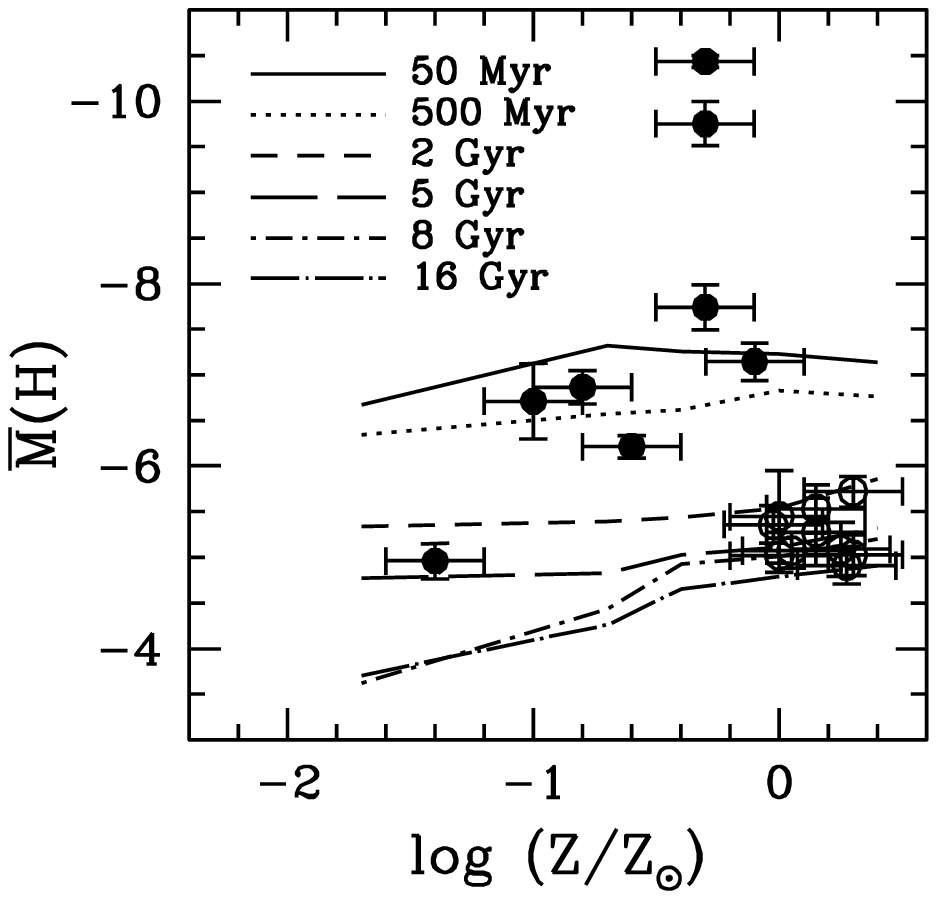}
\includegraphics[clip=,width=0.32\textwidth]{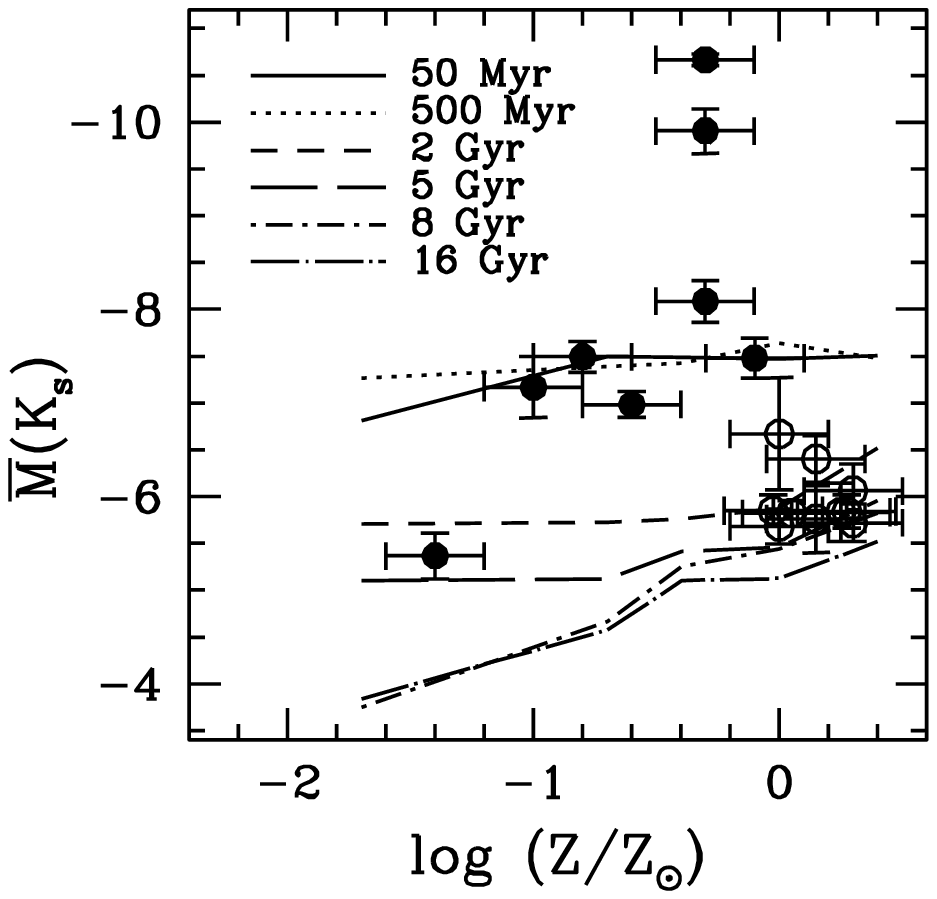}
\caption{Comparison of $J$-band ({\it left}), $H$-band ({\it middle}),
and $K_s$-band ({\it right}) SBF measurements vs. stellar population
metallicity with stellar population synthesis models. All points 
sources with 1\arcmin of the superclusters' centres have been used
to estimate MC SBF magnitudes. Models and symbols are the same as 
in Fig.\ \ref{sbf_mag_age}. }
\label{sbf_mag_z}
\end{figure*}

Fig.\ \ref{sbf_mag_z} shows the evolution of $J$-band, $H$-band, 
and $K_s$-band absolute fluctuation magnitudes respectively for 
MC superclusters and Fornax Cluster galaxies as a function of 
metallicity. Symbols are the same as in Fig.\ \ref{sbf_mag_age}.
MC supercluster fluctuation magnitudes are estimated using all 
point sources within 1\arcmin\ from the centres of the superclusters.  
Different lines connect SSP models with different metallicities, 
but with similar age. The agreement of the predicted relationships 
between near-IR SBF magnitudes and stellar population metallicity 
with the observed ones for MC superclusters, given their age, is 
satisfactory. 
The ages assigned to Fornax Cluster galaxies through their locations 
in the ${\rm M(\barH) \,vs.\,\log(Z/Z_{\odot})}$ diagram, i.e., ages 
older than $\sim\,2$ Gyr, are consistent with the age range of 
Kuntschner (1998). However, few Fornax Cluster galaxies show much 
brighter $K_s$-band SBF magnitudes than those of metal-rich stellar 
populations older than $\sim\,2$ Gyr. The $K_s$-band SBF magnitude 
is probably more sensitive to recent star formation events than 
bluer passbands. 

The SSPs models show, independently of stellar population metallicity, 
that older stellar populations display fainter near-IR fluctuation 
absolute magnitudes. The ability to assign an age to a stellar population, 
given its near-IR absolute fluctuation magnitudes, decreases with age, 
independently of metallicity. The short migration of $K_s$-band 
fluctuation absolute magnitude to brighter magnitudes is clear, i.e., 
the 500~Myr old SSP models are as bright as the 50~Myr old SSP models. 
The figure shows that the sensitivity of near-IR fluctuation magnitudes 
to stellar metallicity, at a given age, increases as one goes from 
$J$-band to $K_s$-band, with fluctuation magnitudes in $J$-band showing
only a weak dependency on metallicity. This behaviour does not depend 
on age, at least in the range investigated in this work; moreover, such 
behaviour is expected. It is suspected that the evolution of SSP model 
properties in a filter around $J$-band is virtually independent of 
metallicity; the evolution of SSP properties reverses there, going from 
optical to near-IR wavelengths (Worthey 1994, Mouhcine \& Lan\c{c}on 2003). 
This suggests that $J$-band fluctuation magnitudes may be used as distance 
indicators, independently of galaxy metallicity. On the other hand, due 
to the sensitivity of $K_s$-band fluctuation magnitudes to metallicity, 
they may be used as a stellar population tracer. Coupled with their high 
luminosity, the predicted evolution of near-IR fluctuation absolute 
magnitudes supports the increasing number of SBF work in these bands. 

\begin{figure}
\includegraphics[clip=,width=0.45\textwidth]{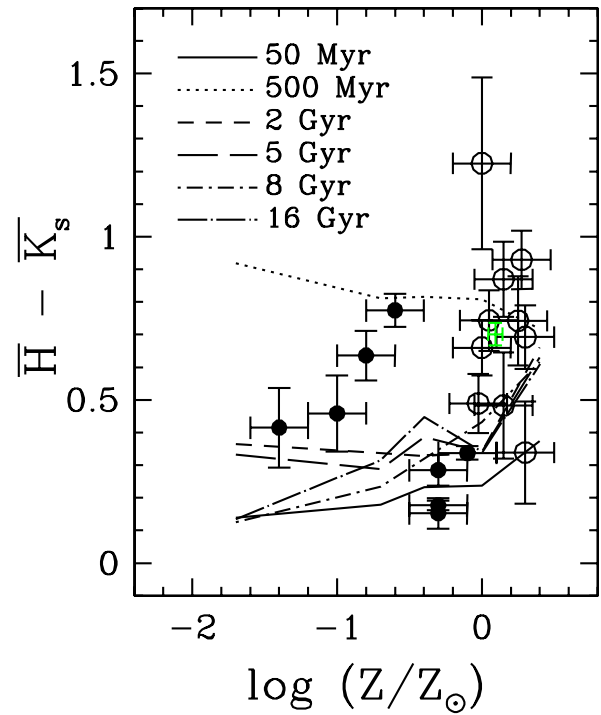}
\caption{Comparison of near-IR fluctuation colour $(\barH-\barKs)$ 
versus stellar population metallicity with the predictions of 
stellar population synthesis models. Models and symbols are akin to 
those in Fig.\ \ref{sbf_mag_age}. All points sources with 1\arcmin 
of the superclusters' centres have been used. The grey error bars 
in the middle of the figure represent the Fornax cluster galaxy 
average.}
\label{sbf_col_z}
\end{figure}

Fig.\ \ref{sbf_col_z} shows a comparison between the observed and 
predicted relationship of $(\barH-\barKs)$ fluctuation colour and 
stellar population metallicity, for MC superclusters and Fornax Cluster 
galaxies. MC supercluster fluctuation colours are estimated using all 
point sources within 1\arcmin\ from the centres of the superclusters.    
Symbols and lines are the same as in Fig.\,\ref{sbf_mag_z}. 
The observed $(\barH-\barKs)$ fluctuation colours for MC superclusters 
agree well the predicted ones given their ages and metallicities.
As discussed in section \ref{sbf_age}, the $(\barH-\barKs)$ fluctuation 
colours estimated using only point source with good photometry are much 
redder that what is predicted given their ages and metallicities. 
In that case, the observed $(\barH-\barKs)$ fluctuation colours for 
the two oldest MC superclusters are compatible with fluctuation colours 
predicted for much younger stellar populations, i.e., between 0.5~Gyr 
and 1.5~Gyr old, dominated by AGB and RGB stars. The discrepancy between 
observed and predicted $(\barH-\barKs)$ fluctuation colour of Fornax 
Cluster galaxies is apparent. Fluctuation colours of a few Fornax galaxies 
are redder than even the reddest models, i.e., these for 500~Myr. 
The ages inferred for these galaxies through this procedure are much 
younger than the ages older than $\sim\,2$ Gyr assigned by Kuntschner 
(2000). This suggests again that the single burst model is not a good 
representation of the star formation histories of all early-type Fornax 
Cluster galaxies.

The evolution of $(\barH-\barKs)$ fluctuation colour versus metallicity 
at a fixed age is complex, and has almost no ability for stellar population 
age-dating. Due to the combined effects of red near-IR colours of carbon 
stars, and the sensitivity of carbon star formation rate to metallicity, 
the fluctuation colour does not simply get redder with metallicity. 
For 500~Myr old SSP models, $(\barH-\barKs)$ 
gets bluer as the metallicity increases, the opposite evolution one may 
expect using a simple opacity argument, due to larger carbon star formation 
efficiency at lower metallicity. This model is significantly redder than
what is predicted by Gonz\'alez et al. (2004) based on Bruzual 
\& Charlot (2003) isochrones, which show redder colours for more 
metal-rich stellar populations. This emphasises again the effects of 
differences in the inclusion of AGB stars in stellar population synthesis 
models. For SSP models where AGB stars are not fueling the near-IR SBF 
signal, the global trend is that young and/or metal-rich stellar 
populations get redder near-IR fluctuation colours than old and/or 
metal-poor ones.

\begin{table}
 \caption{Near-IR SBF predictions from new models}
 \label{models_tab}
 \begin{tabular}{@{}lccccccc}
\hline
   Z & $\log(age)$ & $\bar{J}$ & $\bar{H}$ & $\bar{K_s}$  \\
 \hline
  
 ${\rm Z/Z_{\odot} = 1/50}$ &  7.08  &  -7.541  &  -8.244  &  -8.470  \\   
                            &  7.70  &  -6.047  &  -6.676  &  -6.815  \\  
                            &  8.00  &  -5.460  &  -6.144  &  -6.304  \\  
                            &  8.70  &  -4.788  &  -6.347  &  -7.265  \\  
                            &  9.00  &  -4.715  &  -5.985  &  -6.370  \\   
                            &  9.30  &  -4.192  &  -5.339  &  -5.703  \\   
                            &  9.50  &  -4.031  &  -5.158  &  -5.517  \\   
                            &  9.70  &  -3.734  &  -4.769  &  -5.103  \\  
                            &  9.90  &  -3.022  &  -3.620  &  -3.745  \\  
                            & 10.10  &  -3.046  &  -3.656  &  -3.786  \\   
                            & 10.20  &  -3.077  &  -3.698  &  -3.831  \\  
 \hline 
 ${\rm Z/Z_{\odot} = 1/5}$ &  7.08  &  -8.790  &  -9.460  &  -9.660  \\
                           &  7.70  &  -6.614  &  -7.319  &  -7.499  \\
                           &  8.00  &  -5.554  &  -6.350  &  -6.573  \\
                           &  8.70  &  -5.078  &  -6.574  &  -7.386  \\
                           &  9.00  &  -4.932  &  -6.166  &  -6.535  \\
                           &  9.30  &  -4.264  &  -5.391  &  -5.729  \\
                           &  9.50  &  -4.011  &  -5.106  &  -5.415  \\
                           &  9.70  &  -3.758  &  -4.831  &  -5.119  \\ 
                           &  9.90  &  -3.427  &  -4.433  &  -4.667  \\ 
                           & 10.10  &  -3.411  &  -4.434  &  -4.691  \\  
                           & 10.20  &  -3.303  &  -4.252  &  -4.566  \\ 
  
 \hline

 ${\rm Z/Z_{\odot} = 1/2.5}$  &  7.08  &  -8.869  &  -9.623  &  -9.847  \\  
                              &  7.70  &  -6.444  &  -7.257  &  -7.490  \\   
                              &  8.00  &  -5.428  &  -6.299  &  -6.577  \\   
                              &  8.70  &  -5.110  &  -6.613  &  -7.429  \\   
                              &  9.00  &  -4.683  &  -6.179  &  -7.015  \\    
                              &  9.30  &  -4.311  &  -5.436  &  -5.763  \\    
                              &  9.50  &  -4.104  &  -5.201  &  -5.517  \\    
                              &  9.70  &  -3.854  &  -5.031  &  -5.417  \\    
                              &  9.90  &  -3.840  &  -4.932  &  -5.252  \\    
                              & 10.10  &  -3.631  &  -4.725  &  -5.048  \\    
                              & 10.20  &  -3.479  &  -4.647  &  -5.093  \\ 
  
 \hline

 ${\rm Z/Z_{\odot} = 1}$ &  7.08  &  -9.040  &  -9.740  &  -9.949  \\
                         &  7.70  &  -6.368  &  -7.231  &  -7.469  \\
                         &  8.00  &  -6.174  &  -7.469  &  -8.178  \\ 
                         &  8.70  &  -5.406  &  -6.828  &  -7.635  \\ 
                         &  9.00  &  -4.967  &  -6.345  &  -7.183  \\  
                         &  9.30  &  -4.405  &  -5.535  &  -5.877  \\  
                         &  9.50  &  -4.373  &  -5.552  &  -6.171  \\  
                         &  9.70  &  -4.004  &  -5.114  &  -5.456  \\  
                         &  9.90  &  -3.850  &  -5.010  &  -5.442  \\  
                         & 10.10  &  -3.671  &  -4.771  &  -5.111  \\ 
                         & 10.20  &  -3.682  &  -4.777  &  -5.117  \\   
 \hline
  
 ${\rm Z/Z_{\odot} = 2.5}$ &  7.08  &  -9.012  &  -9.901  & -10.255  \\  
                           &  7.70  &  -6.174  &  -7.135  &  -7.509  \\ 
                           &  8.00  &  -6.122  &  -7.411  &  -8.164  \\ 
                           &  8.70  &  -5.367  &  -6.760  &  -7.477  \\ 
                           &  9.00  &  -5.037  &  -6.393  &  -7.096  \\
                           &  9.30  &  -4.582  &  -5.860  &  -6.521  \\
                           &  9.50  &  -4.314  &  -5.564  &  -6.214  \\ 
                           &  9.70  &  -4.087  &  -5.323  &  -5.953  \\  
                           &  9.90  &  -3.960  &  -5.205  &  -5.821  \\  
                           & 10.10  &  -3.817  &  -5.078  &  -5.694  \\  
                           & 10.20  &  -3.662  &  -4.896  &  -5.507  \\ 
 \hline 
\end{tabular}
\end{table}

\begin{table}
\caption{MC supercluster near-IR SBF magnitudes when all stars
within 1\arcmin\ from the centres of the superclusters are included.}
\label{sbf_tab}
\begin{tabular}{@{}lccccccc}
\hline
Supercluster  & $\bar{J}$ & $\bar{H}$  & $\bar{K_s}$  \\
\hline

Pre-SWB & $-8.93  \pm 0.363$ & $-9.753 \pm 0.239$  & $-9.905 \pm 0.240$ \\
I       & $-9.491 \pm 0.087$ &$-10.43  \pm 0.063$  &$-10.67  \pm 0.061$ \\
II      & $-6.865 \pm 0.313$ & $-7.743 \pm 0.427$  & $-8.087 \pm 0.223$ \\
III     & $-6.145 \pm 0.197$ & $-7.143 \pm 0.207$  & $-7.479 \pm 0.216$ \\
IV      & $-5.772 \pm 0.178$ & $-6.861 \pm 0.183$  & $-7.493 \pm 0.166$ \\
V       & $-4.964 \pm 0.118$ & $-6.213 \pm 0.126$  & $-6.983 \pm 0.138$ \\
VI      & $-5.488 \pm 0.414$ & $-6.712 \pm 0.412$  & $-7.168 \pm 0.326$ \\
VII     & $-4.131 \pm 0.252$ & $-4.959 \pm 0.192$  & $-5.365 \pm 0.245$ \\
\hline
\end{tabular}
\end{table}


\section{Summary \& Conclusions}
\label{concl}

In this paper we have presented new theoretical predictions of 
near-IR SBFs for single age, single metallicity stellar populations. 
The models cover a wide range of metallicities, from 1/50 to 2.5 
times solar, and ages, from 12\,Myr to 16\,Gyr. Our $JHK$ SBF 
predictions are based on the stellar population synthesis models 
of Mouhcine \& Lan\c{c}on (2002), where late-type stellar 
evolutionary phases, expected to dominate the near-IR SBF signal, 
are included with a particular care. The stellar evolution 
prescriptions and stellar libraries used to calculate the stellar 
population isochrones benefit from an extensive comparison with 
various observational constraints. 

The predicted SBF magnitudes and colours have been tested 
against observed near-IR SBFs for MC globular clusters and a 
sample of early-type Fornax Cluster galaxies. The data set covers 
large ranges of ages and metallicities. For MC superclusters 
younger than ${\rm \sim\,3\,Gyr}$, the observed evolution of 
near-IR fluctuation magnitudes and colours as a function of both
stellar population age and metallicity are fairly well reproduced 
by SSP models. 
However, for older MC superclusters the models are not able to 
reproduce the observed near-IR fluctuation absolute magnitudes 
and colours simultaneously. The quality of the agreement between 
the predicted and observed near-IR brightness fluctuations depends 
on how the observational data are treated. 
The discrepancy between the predicted and observed properties of 
old MC superclusters is more likely due to observational reasons. 
Contamination by foreground sources, badly determined photometry 
of stars in the central, most crowded, regions, or even 
overexcision of these same stars could produce the recorded 
disagreement. It would be hence desirable to reobserve MC star 
clusters with SWB types VI and VII, in addition to young star 
clusters, in the near-IR with better spatial resolution and sky 
stability, in order to ascertain whether the origin of the 
discrepancy lies in the models or in the data.
On the other hand, a few Fornax Cluster galaxies display brighter 
near-IR SBF magnitudes and redder fluctuation colours than what 
is predicted given the galaxy ages and metallicities. A possible 
solution of the observed discrepancy is that the single burst 
scenario is not accurate to model the star formation histories 
of these galaxies.

\section*{Acknowledgements}
M. M. would like to thank A. Lan\c{c}on for useful and enlightening
discussions. RAG acknowledges funding from DGAPA--UNAM and CONACYT, 
Mexico. We thank the anonymous referee for a careful reading and for 
requests that improved the paper.

\bsp

\label{lastpage}

\end{document}